\documentclass[12pt]{article}
\usepackage{fancybox}
\usepackage[nosort]{cite}
\usepackage{amsmath,amsfonts,amsbsy}
\usepackage{amssymb,amsthm}
\usepackage{pstricks}
\usepackage{pst-node}
\psset{unit=1cm,linewidth=.5pt,radius=.2}  % controls the size of diagrams

\long\def\symbolfootnote[#1]#2{\begingroup%
\def\thefootnote{\fnsymbol{footnote}}\footnote[#1]{#2}\endgroup}

\def\hybrid{
        \topmargin -20pt
        \oddsidemargin 0pt
        \headheight 0pt \headsep 0pt
        \textwidth 6.25in       % A4 paper
        \textheight 9.5in       % A4 paper
        \marginparwidth .875in
        \parskip 5pt plus 1pt   \jot = 1.5ex}

\hybrid
\newcommand{\normalNode}[2]{\Cnode(#1){#2}}
\newcommand{\dualityNode}[2]{\Cnode[fillstyle=solid,fillcolor=lightgray](#1){#2}}
\newcommand{\disabledNode}[2]{\Cnode[fillstyle=solid,fillcolor=black](#1){#2}}
\newcommand{\singleConnection}[2]{\ncline{-}{#1}{#2}}
\newcommand{\nodeLabel}[2]{\nput[labelsep=0.2]{-40}{#1}{\tiny{#2}}}

\newcommand{\cA}{{\cal A}}
\newcommand{\cB}{{\cal B}}
\newcommand{\cC}{{\cal C}}
\newcommand{\cD}{{\cal D}}
\newcommand{\cE}{{\cal E}}

\newcommand{\cK}{{\cal K}}
\newcommand{\cL}{{\cal L}}
\newcommand{\cM}{{\cal M}}
\newcommand{\cN}{{\cal N}}

\newcommand{\cP}{{\cal P}}
\newcommand{\cQ}{{\cal Q}}

\newcommand{\cV}{{\cal V}}

\newcommand{\beq}{\begin{equation}}
\newcommand{\eeq}{\end{equation}}
\newcommand{\bi}{\begin{itemize}}
\newcommand{\ei}{\end{itemize}}
\newcommand{\bea}{\begin{eqnarray}}
\newcommand{\eea}{\end{eqnarray}}
\newcommand{\ba}{\begin{array}}
\newcommand{\ea}{\end{array}}
\newcommand{\bt}{\begin{tabular}}
\newcommand{\et}{\end{tabular}}
\newcommand{\bc}{\begin{center}}
\newcommand{\ec}{\end{center}}

\def\theequation{\arabic{section}.\arabic{equation}}

\newcommand{\ft}[2]{{\textstyle {\frac{#1}{#2}} }}

\DeclareMathOperator{\ad}{ad}
\newcommand{\g}{\mathfrak{g}}

\newcommand{\R}{\mathbb{R}}

\def\R{\mathbb{R}}

\newcommand\nn{\nonumber}

\newcommand{\pa}{\partial}

\newcommand{\la}{\langle}
\newcommand{\ra}{\rangle}

\def\frake{\mathfrak{e}}

\def\g{\mathfrak{g}}
\def\so{\mathfrak{so}}

\def\sl{\mathfrak{sl}}
\def\gl{\mathfrak{gl}}
\newcommand{\al}{\alpha}
\newcommand{\be}{\beta}
\newcommand{\ep}{\varepsilon}
\newcommand{\de}{\delta}
\newcommand{\ga}{\gamma}
\newcommand{\Ga}{\Gamma}
\newcommand{\Th}{\Theta}
\newcommand{\ttA}{\mathtt{A}}
\newcommand{\ttB}{\mathtt{B}}
\newcommand{\ttC}{\mathtt{C}}

\newcommand{\ttE}{\mathtt{E}}
\newcommand{\ttF}{\mathtt{F}}
\newcommand{\ttG}{\mathtt{G}}
\newcommand{\tth}{\mathtt{h}}

\newcommand{\ttn}{\mathtt{n}}
\newcommand{\ttP}{\mathtt{P}}
\newcommand{\ttQ}{\mathtt{Q}}

\newcommand{\ttg}{\mathtt{g}}
\newcommand{\tte}{\mathtt{e}}
\newcommand{\vA}{\mathcal{A}}
\newcommand{\vB}{\mathcal{B}}
\newcommand{\vC}{\mathcal{C}}
\newcommand{\vD}{\mathcal{D}}
\newcommand{\vE}{\mathcal{E}}
\newcommand{\vF}{\ensuremath{\mathcal{F}}}
\newcommand{\vG}{\ensuremath{\mathcal{G}}}
\newcommand{\vH}{\ensuremath{\mathcal{H}}}

\newcommand{\vL}{\ensuremath{\mathcal{L}}}
\newcommand{\vM}{\ensuremath{\mathcal{M}}}
\newcommand{\vN}{\ensuremath{\mathcal{N}}}
\newcommand{\vP}{\ensuremath{\mathcal{P}}}
\newcommand{\vQ}{\ensuremath{\mathcal{Q}}}
\newcommand{\vR}{\ensuremath{\mathcal{R}}}

\newcommand{\vV}{\ensuremath{\mathcal{V}}}

\newcommand{\coC}{{\frak C}}

\newcommand{\gJ}{\ensuremath{\mathcal{J}}}
\newcommand{\gS}{\ensuremath{\mathcal{S}}}
\newcommand{\gD}{\ensuremath{\mathcal{D}}}

\numberwithin{equation}{section}

\begin{document}

%\begin{titlepage}
\begin{center}

\hfill UG-08-13 \\
\hfill ULB-TH/08-34 \\
\hfill AEI-2008-083 \\

\vskip 1.5cm

{\Large \bf
$E_{10}$ and gauged maximal supergravity\\[0.2cm]}

\vskip 1.5cm

{\bf%\large
Eric A. Bergshoeff\,\footnotemark[1], Olaf Hohm\,\footnotemark[1], Axel Kleinschmidt\,\footnotemark[2],
\\
\vspace*{1mm}
%\vskip 0.1pt
Hermann Nicolai\,\footnotemark[3],
Teake A. Nutma\,\footnotemark[1], Jakob Palmkvist\,\footnotemark[3]\,\footnotemark[4]} \\

\vskip 15pt

{\em \footnotemark[1]Centre for Theoretical Physics, University of Groningen, \\
Nijenborgh 4, NL-9747 AG Groningen, The Netherlands \vskip 10pt }

{\em \footnotemark[2]Physique Th\'eorique et Mathematique \& International Solvay Institutes,\\
Universit\'e Libre de Bruxelles,\\
Boulevard du Triomphe, ULB-CP 231, BE-1050 Bruxelles, Belgium \vskip 10pt }

{\em \footnotemark[3]Max-Planck-Institut f\"ur Gravitationsphysik, Albert-Einstein-Institut, \\
Am M\"uhlenberg 1, DE-14476 Golm, Germany \vskip 10pt }

\footnotemark[4]{\em  Fundamental Physics, Chalmers University of Technology,\\
 SE-412 96 G\"oteborg, Sweden.}

\vskip 10pt

{\footnotesize{\tt E.A.Bergshoeff@rug.nl, O.Hohm@rug.nl, Axel.Kleinschmidt@ulb.ac.be,}} \\
{\footnotesize{\tt Hermann.Nicolai@aei.mpg.de, T.A.Nutma@rug.nl, Jakob.Palmkvist@aei.mpg.de}}

\vskip 0.8cm

\end{center}

\vskip 1cm

\begin{center} {\bf ABSTRACT}\\[3ex]

\begin{minipage}{13cm}
\small We compare  the dynamics of maximal three-dimensional
gauged supergravity in appropriate truncations with the equations of motion
that follow from a one-dimensional $E_{10}/K(E_{10})$ coset model at the first
few levels. The constant embedding tensor, which describes gauge
deformations and also constitutes an M-theoretic degree of freedom beyond
eleven-dimensional supergravity, arises naturally as an integration constant 
of the  geodesic model. In a detailed analysis, we find complete agreement 
at the lowest levels. At higher levels there appear mismatches, as in previous 
studies. We discuss  the origin  of these mismatches.

\end{minipage}

\end{center}

\noindent

\vfill

October 2008

\thispagestyle{empty}
%\end{titlepage}

\newpage

\tableofcontents

\section{Introduction}%\setcounter{equation}{0}
It is well-known that the highest space-time dimension that allows a
supergravity theory is eleven \cite{Nahm:1977tg}.
Upon a torus reduction to lower dimensions, eleven-dimensional
supergravity \cite{Cremmer:1978km} leads, in each space-time dimension
$3\le D\le 10$, to a maximal supergravity theory in which the scalars
parametrize a coset manifold $G/K(G)$, where $K(G)$ is the maximal compact
subgroup of $G$ \cite{1979NuPhB.159..141C}. For maximal supergravity in
$D=3$ dimensions, the rigid symmetry group is the non-compact split
real form of the largest 
exceptional Lie group $E_8$; all physical bosonic degrees of freedom
reside in the coset space, with no propagating gravitational degrees of
freedom left. This theory was already constructed long ago
\cite{1983uft..conf..215J,Marcus:1983hb}; however, its gauged versions,
whose relation with the infinite-dimensional $E_{10}/K(E_{10})$
coset model will be the focus of the present paper, were obtained
only much more recently \cite{Nicolai:2000sc,Nicolai:2001sv}.

The different duality groups $G$ characterizing the coset manifolds
are described by Dynkin diagrams that are related to each other by
deleting nodes (going up in dimension) or adding nodes (going down
in dimension). The three-dimensional case corresponds to the group
$G=E_8$ which has a Dynkin diagram with $8$ nodes. It has been
suggested that by reducing to even lower dimensions, $0 \le D \le
2$, larger symmetry algebras may emerge that correspond to Dynkin
diagrams which are obtained by adding nodes to the $E_8$ diagram~\cite{Julia:1982gx}.
Such diagrams do not correspond to a finite number of symmetries, as
in the case of ordinary Lie groups, but instead lead to an infinite
number of symmetries corresponding to the infinite-dimensional
groups $E_9$\ ($D=2$), $E_{10}$\ ($D=1$) and $E_{11}$\ ($D=0$),
respectively.

It has been conjectured that maximal supergravity in any dimension
$D \le 11$, independent of any torus reduction, can be described in
terms of $E_{11}$
\cite{West:2001as,Schnakenburg:2001ya,Kleinschmidt:2003mf}. While
this conjecture works well (at low levels) as far as the kinematics
is concerned, yielding the correct bosonic multiplets of various
maximal supergravities upon decomposition of $E_{11}$ under its
finite-dimensional subalgebras, the underlying dynamics is much less
understood. In this paper, we will therefore follow a different route, based
on a conjecture proposed and elaborated in
\cite{Damour:2002cu,Damour:2004zy}, according to which the dynamics
of any maximal supergravity theory (or some M-theoretic extension
thereof) is described by the equations of motion of a
one-dimensional sigma model over the coset space $E_{10}/K(E_{10})$.
If these equations are supplemented by coset
constraints~\cite{Damour:2007dt}, one can establish a correspondence
between truncated versions of the coset equations on the one hand,
and of the supergravity equations on the other. This correspondence
can also be extended to the fermionic sector such that the fermionic
field equations can be reformulated to be covariant under the coset
model `R symmetry'
$K(E_{10})$~\cite{Damour:2005zs,deBuyl:2005mt,Damour:2006xu}.\footnote{An
  approach combining ideas of the $E_{10}$ and $E_{11}$ approaches has been
  explored in \cite{Englert:2003py,Englert:2004it}.} 

For carrying out the comparison one has to formulate both sides of the
correspondence appropriately.
On the one hand one
has to truncate the supergravity fields
and break space-time
covariance by choosing an ADM gauge, in order to be amenable to a
one-dimensional language. On the $E_{10}$ side, on the other hand,
one has to perform a so-called level decomposition with respect to
the subgroup $GL(D-1)\times G_{D}$, where $G_D$ denotes the duality
group in $D$ dimensions. At low levels, the equations of motion of
the $E_{10}$ model precisely match the equations of motion of (pure)
supergravity truncated to only a time-dependent, that is,
one-dimensional system. This matching is in accord with the
(duality) symmetries expected to appear in lower dimensions.
However, the main challenge is to go beyond these low levels and to
find an interpretation for the infinite tower of representations
appearing in the level decomposition of $E_{10}$ and $E_{11}$ (see
e.g. \cite{Nicolai:2003fw,Kleinschmidt:2003mf}) also on the
supergravity side.

As one attractive scenario it has been suggested
\cite{Damour:2002cu,Nicolai:2003fw,Damour:2004zy} that the higher
levels encode the spatial gradients of the supergravity fields, and
so by including all of these states one should finally recover the
full unrestricted supergravity in $D$ dimensions or an M-theoretic extension
thereof.\,\footnote{In the $E_{11}$ approach some of the higher level
states can be interpreted as dual representations of lower
level states \cite{Riccioni:2006az}.} While some intriguing confirmation
has been found, certain mismatches remain, such that a conclusive
picture of how to identify the spatial dependence within $E_{10}$ 
and how to understand the emergence of a space-time field theory from the
one-dimensional sigma model is still lacking.

Another interpretation for part of the higher levels concerns
certain mass deformations of pure maximal supergravity. In
\cite{Kleinschmidt:2004dy} it has been shown that the massive Romans
supergravity in ten dimensions \cite{Romans:1985tz}, which deforms
type IIA supergravity by a mass parameter $m$, is contained in the
$E_{10}$ model, upon taking a certain $9$-form representation into
account (see also~\cite{Henneaux}). For the realization of massive type IIA
supergravity within  
the $E_{11}$ approach see \cite{Schnakenburg:2002xx}.

Apart from switching on spatial gradients and/or mass parameters,
another direction will be explored in this paper, namely that of
turning on gauge couplings. This possibility relies on the recent
realization that $E_{11}$ and $E_{10}$ contain information about
gauged supergravity via $D$- and $(D-1)$-form representations
\cite{Riccioni:2007au,Bergshoeff:2007qi,Bergshoeff:2007vb,Riccioni:2007ni}.\footnote{The
$D$-form representations only occur in the $E_{11}$ approach.} We
will focus on gauged supergravity in three dimensions, but our
conclusions are expected to be of general validity. The advantage of
this case is that $E_{8}$ is the largest finite-dimensional duality
group. As a consequence, the $E_{10}$ equations of motion truncated
to level $\ell =0$ already match ungauged supergravity reduced to a
one-dimensional system. Thus, this model allows a clear distinction
between the `manifest' aspects of the $E_{10}$ conjecture at level
$\ell=0$ and the more speculative features related to higher levels,
as spatial gradients or gauge couplings. We will find surprising
correpondences between both sides, but also mismatches, which remain
to be investigated further.

Let us emphasize the main features of our results, also reflecting the
differences with the $E_{11}$ approach
\cite{West:2001as,Riccioni:2007au,Riccioni:2007ni}. These are:
\begin{itemize}
\item There is {\em no need to deform the $E_{10}$ Lie algebra}
   or the $E_{10}$ Cartan form (e.g. by modifying the derivative)
   in order to obtain agreement (as far as it goes) between the equations
   of gauged $D=3$ supergravity and the $E_{10}/K(E_{10})$ coset
   model. Rather, the gauging appears exclusively as a consequence of
   `switching on' certain higher level degrees of freedom in the
   level expansion of the Cartan form and the coset equations of motion.
   The relevant components of the embedding tensor are in part beyond
   level $\ell=3$ in the $SL(10)$ decomposition, hence cannot be understood
   via Kaluza-Klein-type compactification from $D=11$ supergravity (as also emphasized in
   \cite{Riccioni:2007au}).
\item The absence of any deformation in the original coset model, in turn,
   is a direct consequence of the fact that the correspondence works only
   if we adopt the {\em temporal gauge} for all gauge fields, and in
   particular for the Chern-Simons gauge potential $\ttA_\mu{}^\cM$ (generalizing
   the pseudo-Gaussian gauge, i.e. vanishing shift, for the gravitational
   degrees of freedom).
\item We are here working in a  Hamiltonian framework. This means that
   in addition to the coset equations of motion (which are related to
   the evolution equations involving time derivatives on the supergravity
   side) we need to impose certain canonical constraints on the coset
   dynamics (corresponding to constraints on the initial data on the
   supergravity side). The structure of these constraints was
   studied in~\cite{Damour:2007dt}, and we here likewise find that the constraints can be
   written in a Sugawara-like form in terms of the coset variables. One
   can also show that under (part of) $E_{10}$ the constraints
   transform into one another, such that duality relates for instance
   the diffeomorphism constraint and the quadratic constraint of gauged
   supergravity. This feature is somewhat reminiscent of the $L(\Lambda_1)$
   representation found in \cite{Riccioni:2007ni}, but the precise relation
   (if any) is not clear (e.g. in~\cite{Damour:2007dt} the constraints were found {\em not}
   to transform as a highest or lowest weight representation of the whole $E_{10}$).
\end{itemize}

The paper is organized as follows. In section 2 we first summarize
the $E_{10}/K(E_{10})$ coset model. In particular, we derive the
equations of motion at the lowest levels. In section 3 we consider
maximal gauged supergravity in three dimensions and its torus
reduction to one (time) dimension. Next, in section 4 we discuss the
supergravity/$E_{10}$ correspondence: its matches and mismatches.
Finally, in section 5 we give our outlook on the status of the
$E_{10}$ conjecture. We include two appendices summarizing some basic
properties of $E_8$ and the details about the level decomposition of $E_{10}$.

\section{The $E_{10}/K(E_{10})$ coset model}

In this section we introduce the $E_{10}/K(E_{10})$ coset model. In order to make contact with three-dimensional gauged supergravity it proves convenient to write the generators of $E_{10}$ in a $SL(2,\mathbb{R})\times E_{8(8)}$ covariant form. We then analyze the one-dimensional coset model in this language and derive the associated geodesic equations.

By $\frake_8$ and $\frake_{10}$ we always
mean the split real forms
(also denoted $\frake_{8(8)}$ and $\frake_{10(10)}$)
of the corresponding complex Lie algebras. The Lie groups obtained by exponentiation of the algebra elements are denoted $E_8$ and $E_{10}$.
Sometimes the notation $\frake_8{}^{++}$ and $E_{8}{}^{++}$ is used, indicating that $\frake_{10}$ is the `over-extension' of $\frake_8$ -- the Dynkin diagram of $\frake_{10}$ is obtained by adding two extra nodes to that of $\frake_{8}$, as can be seen from Figure \ref{e10dynkin1}.

\subsection{Generalities about $E_{10}$} \label{generalities}
\begin{figure}[h]
\begin{center}
\begin{pspicture}(0,0)(8,1)
\normalNode{6,1}{N12062844712}\nodeLabel{N12062844712}{10}
\normalNode{0,0}{N22062844712}\nodeLabel{N22062844712}{1}
\normalNode{1,0}{N32062844712}\nodeLabel{N32062844712}{2}
\normalNode{2,0}{N42062844712}\nodeLabel{N42062844712}{3}
\normalNode{3,0}{N52062844712}\nodeLabel{N52062844712}{4}
\normalNode{4,0}{N62062844712}\nodeLabel{N62062844712}{5}
\normalNode{5,0}{N72062844712}\nodeLabel{N72062844712}{6}
\normalNode{6,0}{N82062844712}\nodeLabel{N82062844712}{7}
\normalNode{7,0}{N92062844712}\nodeLabel{N92062844712}{8}
\normalNode{8,0}{N102062844712}\nodeLabel{N102062844712}{9}
\singleConnection{N22062844712}{N32062844712}
\singleConnection{N32062844712}{N42062844712}
\singleConnection{N42062844712}{N52062844712}
\singleConnection{N52062844712}{N62062844712}
\singleConnection{N62062844712}{N72062844712}
\singleConnection{N72062844712}{N82062844712}
\singleConnection{N82062844712}{N92062844712}
\singleConnection{N92062844712}{N102062844712}
\singleConnection{N12062844712}{N82062844712}
\end{pspicture}
\end{center}
\caption{The Dynkin diagram of $E_{10} = E_{8}{}^{++}$}\label{e10dynkin1}
\end{figure}
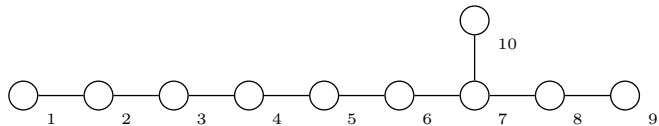

We first briefly summarize some basic facts about $E_{10}$. Its Lie algebra
is characterized by the Dynkin diagram given in Figure~\ref{e10dynkin1}.

\noindent
More precisely, the Lie algebra $\frak{e}_{10}$ of $E_{10}$ is
defined in terms of a $10\times 10$ Cartan matrix $A_{ij}\ (i,\,j = 1,\,
\ldots,\,10)$, which can be read off from the Dynkin diagram as
 \bea
    A_{ij} =
        \begin{cases}
            2   & \text{if $i=j$}, \\
            -1  & \text{if there is a line between nodes $i$ and $j$}, \\
            0   & \text{otherwise}.
        \end{cases}
\eea
The Lie algebra is then generated by multiple commutators of the ten
basic triples of generators $\{h_i, e_i, f_i\}$. %_{i = 1,\ldots,r}$,
%where $r$ is the rank of the Cartan matrix.
The $h_i$ are elements
of the abelian Cartan subalgebra. The
%, satisfying $[h_i, h_j]  = 0$. The
$e_i$ and $f_i$ are the positive and negative step operators. Their
commutation relations (the \textit{Chevalley relations}) read
\bea \label{chevrel}
    [h_i, e_j]  = A_{ij} e_j \;, \quad
    [h_i, f_j]  = -A_{ij} f_j\;, \quad
    [e_i, f_j]  = \delta_{ij} h_i\;
\eea
(no summation).
The multiple commutators
are constrained by the \textit{Serre relations} \bea \label{serrerel}
    (\ad_{e_i})^{1-A_{ij}} e_j  = 0\;, \qquad
    (\ad_{f_i})^{1-A_{ij}} f_j  = 0\;.
\eea

Each Kac-Moody algebra admits an invariant Cartan-Killing form, which in the
basis introduced above reads
 \bea \label{ckform}
  \la e_i|f_j \ra \ = \ \delta_{ij}\;, \qquad
  \la h_i|h_j \ra \ = \ A_{ij}\;.
 \eea
We note that the Cartan matrix $A_{ij}$, and thereby the Cartan-Killing form on the Cartan subalgebra, is of Lorentzian signature.
This will later be used to define a
\textit{null-geodesic} motion on the coset space $E_{10}/K(E_{10})$.
We also need the Chevalley involution $\omega$ in order to
define the maximal compact subgroup $K(E_{10})$ and its Lie
algebra $\mathfrak{k}(\frake_{10})$.
The Chevalley involution is
defined by \bea
    \omega(e_i)   = - f_i \;, \quad
    \omega(f_i)  = - e_i \;, \quad
    \omega(h_i)  = - h_i \;.
\eea
One then defines the (generalized) transpose of an $\frake_{10}$ element $x$ as $x^T=-\omega(x)$.
The maximal compact subalgebra $\mathfrak{k}(\frake_{10})$ is
defined as the subalgebra of $\frake_{10}$ that is pointwise fixed
by the Chevalley involution. Thus it consists of all elements
$x-x^T$. Similarly, we define the coset
$\frake_{10}\ominus\mathfrak{k}(\frake_{10})$ to be the subspace
consisting of all elements $x+x^T$. With respect to the Cartan-Killing form, the maximal compact subalgebra
$\mathfrak{k}(\frake_{10})$ is negative-definite, the coset
$\frake_{10} \ominus
 \mathfrak{k}(\frake_{10})$ 
 is almost positive-definite (there is one negative eigenvalue of the
 Cartan-Killing metric in the Cartan subalgebra),
and these two subspaces of $\frake_{10}$
are orthogonal complements to each other.

\subsection{Decomposition under $SL(2,\,\mathbb{R}) \times E_{8}$}
\label{decompsection}

Any Kac-Moody algebra can be written as a
direct sum of subspaces
$\g_\ell$ for all integers $\ell$ such that
\begin{align} \label{gradstrukt}
[\g_{k},\,\g_{\ell}] \subseteq \g_{k + \ell}.
\end{align}
For $k=0$, this gives a level decomposition of the adjoint representation of $\frake_{10}$ under a subalgebra $\g_0$, where we
call $\ell$ the \textit{level} of the elements in $\g_\ell$, and of the corresponding
$\g_0$
representation.

\begin{figure}[h] 
\begin{center}
\begin{pspicture}(0,0)(8,1)
\dualityNode{6,1}{N11888086120}\nodeLabel{N11888086120}{10}
\normalNode{0,0}{N21888086120}\nodeLabel{N21888086120}{1}
\disabledNode{1,0}{N31888086120}\nodeLabel{N31888086120}{2}
\dualityNode{2,0}{N41888086120}\nodeLabel{N41888086120}{3}
\dualityNode{3,0}{N51888086120}\nodeLabel{N51888086120}{4}
\dualityNode{4,0}{N61888086120}\nodeLabel{N61888086120}{5}
\dualityNode{5,0}{N71888086120}\nodeLabel{N71888086120}{6}
\dualityNode{6,0}{N81888086120}\nodeLabel{N81888086120}{7}
\dualityNode{7,0}{N91888086120}\nodeLabel{N91888086120}{8}
\dualityNode{8,0}{N101888086120}\nodeLabel{N101888086120}{9}
\singleConnection{N21888086120}{N31888086120}
\singleConnection{N31888086120}{N41888086120}
\singleConnection{N41888086120}{N51888086120}
\singleConnection{N51888086120}{N61888086120}
\singleConnection{N61888086120}{N71888086120}
\singleConnection{N71888086120}{N81888086120}
\singleConnection{N81888086120}{N91888086120}
\singleConnection{N91888086120}{N101888086120}
\singleConnection{N11888086120}{N81888086120}
\end{pspicture}
\end{center}
\caption{\label{e10dynkin}
    Level decomposition of $E_{10} = E_{8}{}^{++}$.
    The grey nodes denote the duality group $E_{8}$,
    the black node is the deleted one and the white node
    denotes the $SL(2,\R)$ spacetime subgroup.}
\end{figure}
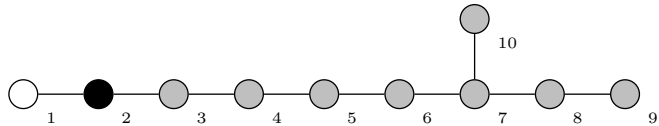

In order to make contact with three-dimensional supergravity we
perform
a level decomposition of $E_{10}$ with
respect to the subgroup of spatial diffeomorphisms and the duality
group:
 \bea
  E_{10} \supset SL(2,\mathbb{R})\times E_{8}\;.
 \eea
This corresponds to deleting the black node numbered 2 in the Dynkin diagram
in figure~\ref{e10dynkin}. 
\vskip12pt

%\newpage

Thus we consider the case where $\g_0 = \gl(2,\,\R) \oplus
\frake_8$, where the enhancement from $\sl(2,\,\R)$ to $ \gl(2,\,\R)$ is due
to the Cartan generator associated with the deleted node~$2$. The
representations occurring in this level decomposition 
can be calculated using the computer program  SimpLie \cite{SimpLie}.
Up to level $\ell < 3$ we find the $\frak{sl}(2,\,\R)\oplus
\frak{e}_{8}$ representations in Table \ref{reptable}, where we
indicated the corresponding generators with their symmetries. We
denote by $a,\,b=1,\,2$ the fundamental indices of $GL(2,\,\R)$ and
by $\cA,\,\cB = 1,\,2\ldots ,248$ the adjoint indices of $E_{8}$.
The fields associated to the $\ell=0$ generators are the spatial zweibein and
the coset scalars. The $\ell=1$ fields can be interpreted as gauge
vectors. The interpretation of the $\ell=2$ fields will be discussed
in section~4.3 (concerning the embedding tensor components $\theta$ and
${\tilde \Theta}$), where also some 
speculations will be made on trombone gaugings. At the
negative levels we have the conjugate representations, i.e., the
transposed generators of those at the positive levels.

\vskip8pt
\begin{table}[h]
\begin{centering}
\begin{tabular}{|c|c|c|c|}
    \hline
    Level $\ell$    & $SL(2,\,\R) \times E_8$ representation    & Generator                         & Interpretation \\
    \hline
    \hline
    &&& \\ [-10pt]
    0               & $( {\bf 1} \oplus {\bf 3}, {\bf 1})$      & $K^a{}_{b}$
    & spatial zweibein \\
    &&& \\ [-10pt]
                    & $( {\bf 1}, {\bf 248})$                   & $t^{\cA}$                         & scalars \\
    &&& \\ [-10pt]
    \hline
    &&& \\ [-10pt]
    1               & $( {\bf 2}, {\bf 248})$                   & $E^a{}_{\cA}$                     & gauge vectors\\
    &&& \\ [-10pt]
    \hline
    &&& \\ [-10pt]
    2               & $( {\bf 1}, {\bf 1})$                     & $E$                               & $\theta$ \\
                    & $({\bf 1}, {\bf 3875})$                   & $E_{\vA\vB}=E_{(\vA\vB)}$         & $\tilde{\Theta}_{\cM\cN}$\\
    &&& \\ [-10pt]
                    & $( {\bf 3}, {\bf 248})$                   & $E^{ab}{}_{\vA}=E^{(ab)}{}_{\vA}$ & trombone gauging? \\  [2pt]
    \hline
\end{tabular}
\caption{
    $SL(2,\,\R) \times E_8$ representations within $E_{10}$ up to level 2.}
    \label{reptable}
\end{centering}
\end{table}

Later we will split the $E_8$ indices as
\begin{align} \label{splitting}
\vA \ \rightarrow \ [IJ],\ A,
\end{align}
where $I,\,J=1,\,2,\,\ldots,\,16$ and $A=1,\,2,\,\ldots,\,128$
are vector and spinor indices, respectively, of the maximal compact subalgebra $\mathfrak{k}(\mathfrak{e}_8)=\so(16)$.
This is in accordance with the following decomposition of the adjoint
$\frake_8$ representation under the $\so(16)$ subalgebra 
\begin{align}
\bf248 \rightarrow \bf120+128\,.
\end{align}

As indicated in Table \ref{reptable}, the generator $E_{\vA\vB}$ is symmetric in the two adjoint $E_8$ indices.
However, it also has to satisfy further conditions in order to belong to the {\bf 3875} representation; in particular
it must be traceless. The
necessary and sufficient condition for this can be expressed as
\begin{align}
\mathbb{P}_{\vA\vB}{}^{\vC\vD}E_{\vC\vD}=E_{\vA\vB}, \label{villkor}
\end{align}
where the explicit form of the projector $\mathbb{P}_{\vA\vB}{}^{\vC\vD}$ has been determined in \cite{Koepsell:1999uj} and reads
\begin{align} \label{komponentform}
\mathbb{P}_{\vA\vB}{}^{\vC\vD} &= \tfrac{1}{7}\delta_{(\vA}{}^{\vC}\delta_{\vB)}{}^{\vD}
-\tfrac{1}{56}\eta_{\vA\vB}\eta^{\vC\vD}-\tfrac{1}{14}f^{\vE}{}_{\vA}{}^{(\vC}f_{\vE\vB}{}^{\vD)}.
\end{align}
Here $f$ and $\eta$ denote the $E_8$ structure constants and the
components of the Killing form, respectively. These are given
explicitly in appendix \ref{e8app}.

At level $\ell=0$ we find
a singlet plus the adjoint of $\frak{sl}(2,\,\R)\oplus \frak{e}_8$.
The first part, $(\bf{1}\oplus \bf{3}, \bf{1})$, can be seen as the adjoint of
$\frak{gl}(2,\,\R)$. The
$\ell =0$ subalgebra reads
\begin{align}\label{l0algebra}
[t^\vA,\,t^\vB]&=f^{\vA\vB}{}_{\vC}t^{\vC},&
[K^a{}_b,\,K^c{}_d]&=\delta^{c}{}_b K^{a}{}_d
-\delta^{a}{}_d K^{c}{}_b. %\nn
\end{align}

The Lie brackets that do not mix between positive and negative
levels are entirely fixed by representation theory and the graded
structure (\ref{gradstrukt}). The commutators involving the $\ell=0$
generators just give the transformation character of the
$|\ell|=1,2$ generators under $\frak{gl}(2,\,\R)\oplus
\frak{e}_{8}$. Since the generators at the negative levels transform
in the conjugate representations compared to the positive levels,
they have their $\sl(2,\,\R)$ indices downstairs instead. However,
the position of the $E_8$ indices is arbitrary in the definition of
the generators, since they can be raised and lowered by means of
the $\frake_8$ Killing form $\eta$, which we describe in
(\ref{Cartan-Killing}). We here define the generators on the
negative levels by the following action of the Chevalley involution:
\begin{align} \label{chevinv1}
\omega(E^a{}_{\vA}) &= -F_a{}^{\vA}
\end{align}
at level $\ell=-1$ and
\begin{align} \label{chevinv2}
\omega(E_{\vA\vB}) &= -F^{\vA\vB}, &  \omega(E) &= -F, & \omega(E^{ab}{}_{\vA}) &= -F_{ab}{}^{\vA}
\end{align}
at level $\ell=-2$. We recall that the transpose then is defined
as $x^T=-\omega(x)$.

The commutators involving level zero are now given by
\begin{align}\label{liealgebra}
[t^\vA,\,E^{a}{}_{\vB}]&=f^{\vA}{}_{\vB}{}^{\vC}E^{a}{}_{\vC},&
[t^\vA,\,F_{a}{}^{\vB}]&=f^{\vA\vB}{}_{\vC}F_{a}{}^{\vC},\nn\\
[K^a{}_b,\,E^{c}{}_{\vA}]&=\delta^c{}_bE^{a}{}_{\vA},&
[K^a{}_b,\,F_{c}{}^{\vA}]&=-\delta^a{}_cF_{b}{}^{\vA},\nn\\
[t^\vA,\,E_{\vB\vC}]&=2f^{\vA}{}_{\vB}{}^{\vD}E_{\vC\vD},&
[t^\vA,\,F^{\vB\vC}]&=2f^{\vA\vB}{}_{\vD}F^{\vC\vD},\nn\\
[t^\vA,\,E^{cd}{}_{\vB}]&=f^{\vA}{}_{\vB}{}^{\vD}E^{cd}{}_{\vD},&
[t^\vA,\,F_{cd}{}^{\vB}]&=f^{\vA\vB}{}_{\vD}F_{cd}{}^{\vD},\nn\\
[K^a{}_b,\,E]&=\delta^a{}_bE,&
[K^a{}_b,\,F]&=-\delta^a{}_bF,\nn\\
[K^a{}_b,\,E_{\vA\vB}]&=\delta^a{}_bE_{\vA\vB},&
[K^a{}_b,\,F^{\vA\vB}]&=-\delta^a{}_bF^{\vA\vB},\nn\\
[K^a{}_b,\,E^{cd}{}_{\vA}]&=2\delta^c{}_bE^{ad}{}_{\vA},&
[K^a{}_b,\,F_{cd}{}^{\vA}]&=-2\delta^a{}_cF_{bd}{}^{\vA}.%\nn
\end{align}
\textit{Here and troughout this paper, we use the convention of implicit (anti-)symmetrization in indices}. This means that the right hand side of any equation is always assumed to be (anti-)symmetrized according to the left hand side. In (\ref{liealgebra})
this convention concerns the generators $E_{\vA\vB}$ and $E^{ab}{}_\vA$ at level $\ell=2$ (and their transposes at level $\ell=-2$), which are symmetric in the $E_8$ and $SL(2,\,\R)$ indices, respectively (cf.~table \ref{reptable}). For example, the last equation in
(\ref{liealgebra}) should be  read as
\begin{align}
 [K^a{}_b,\,F_{cd}{}^{\vA}]&=-\delta^a{}_cF_{bd}{}^{\vA}
 -\delta^a{}_dF_{bc}{}^{\vA}.%\nn
\end{align}
Later, when we split the $E_8$ indices
as in (\ref{splitting}),
this convention will also concern antisymmetric pairs $[IJ]$ of $SO(16)$ vector indices.

We define the generators at level $|\ell|=2$ by the commutation relations
\begin{align}
[E^{a}{}_{\vA},\,E^{b}{}_{\vB}] &= \tfrac12\varepsilon^{ab}\eta_{\vA\vB}E
+\varepsilon^{ab}E_{\vA\vB}-f_{\vA\vB}{}^{\vC}E^{ab}{}_{\vC},\nn\\
[F_{a}{}^{\vA},\,F_{b}{}^{\vB}] &= -\tfrac12\varepsilon_{ab}\eta^{\vA\vB}F
-\varepsilon_{ab}F^{\vA\vB}-f^{\vA\vB}{}_{\vC}F_{ab}{}^{\vC}. \label{postulat1}
\end{align}
We will see below that this normalization is
a convenient choice.
Note that both equations have a minus sign on the last term, but otherwise opposite signs on the right hand side. This is necessary if we want $F^{\vA\vB}$ to be the transpose of $E_{\vA\vB}$, that is, if we want to obtain (\ref{chevinv2}) from (\ref{chevinv1}) using the homomorphism property of $\omega$.
The reason is that
$f^{\vA\vB}{}_{\vC}=-f_{\vA\vB}{}^{\vC}$ for the $\mathfrak{e}_8$ structure constants (see appendix \ref{e8app}), whereas $\eta^{\vA\vB}=\eta_{\vA\vB}$ and
$\de^\vA{}_\vC \de^\vB{}_\vD=\de_\vA{}^\vC \de_\vB{}^\vD$.

As we show in appendix \ref{e10app}
the Chevalley-Serre relations (\ref{chevrel}) and (\ref{serrerel})
lead to
\begin{align}
[E^{a}{}_{\vA},\,F_{b}{}^{\vB}]&=\delta^a{}_bf_{\vA}{}^{\vB}{}_{\vC}t^{\vC}
+\delta_{\vA}{}^{\vB}K^a{}_b-\delta_{\vA}{}^{\vB}\delta^a{}_b K, \label{postulat3}
\end{align}
where we have set
\begin{align}
K=K^a{}_a=K^1{}_1+K^2{}_2.
\end{align}
The remaining non-zero commutation relations up to level $|\ell|=2$ can be
derived from those above by the Jacobi identity. For completeness they are
also given in appendix \ref{e10app}.

We must define the Cartan-Killing form for the generators at level $|\ell| \leq 2$ in a way such that (\ref{ckform}) is satisfied after
identifying the generators in the Chevalley basis (see appendix
\ref{e10app}). This is achieved by the following normalization at
level zero:
\begin{align}
\la K^a{}_b | K^c{}_d \ra&=\delta^a{}_d \delta^c{}_b-\delta^a{}_b \delta^c{}_d,
& \la t^\vA | t^\vB \ra &= \eta^{\vA\vB}, & \la K^a{}_b | t^\vA \ra =0,
\end{align}
which gives back the Cartan-Killing form for 
$\frake_8$. 
For the levels $|\ell|=1,\,2$ %of $E_{10}$
we now get
\begin{align}
\la E^a{}_\vA | F_b{}^\vB \ra &= \delta^a{}_b\delta_{\vA}{}^{\vB}, &
\label{bilform}%\nn%\\
\la E_{\vA\vB} | F^{\vC\vD} \ra &={14}\, \mathbb{P}_{\vA\vB}{}^{\vC\vD},\nn\\%\nn%\\
\la E|F\ra &= 1, &
\la E^{ab}{}_\vA | F_{cd}{}^\vB \ra &= %\label{Cartan2}
\delta^{a}{}_{c}\delta^{b}{}_{d}\delta_{\vA}{}^{\vB}, %\label{bilform2}
\end{align}
and zero elsewhere, using the invariance of the bilinear form.

Taking $x$ to be a basis element of $\frake_{10}$
in the expressions
$x-x^T$ and $x+x^T$, we
obtain bases of $\mathfrak{k}(\frake_{10})$ and the coset
$\frake_{10}\ominus\mathfrak{k}(\frake_{10})$, respectively.
On the $\frake_8$ subalgebra the Chevalley involution acts as
$\omega(t^\vA)=-t_\vA = -\eta_{\vA\vB}t^\vB$. The transpose is
then given by
\begin{align}
(t_\vA)^T = t^\vA.
\end{align}
 On the $\sl(2,\,\R)$ subalgebra, the transpose is just the ordinary transpose,
\begin{align}
(K^a{}_b)^T = K^b{}_a.
\end{align}
 Thus at level zero we define
\begin{align}
\gJ_{IJ} &= t_{IJ}-t^{IJ}=-2t^{IJ}, & \gJ^{ab}&= K^a{}_b - K^b{}_a
\end{align}
  as basis elements of $\mathfrak{k}(\frake_{10})=\mathfrak{so}(16)$ and
  $\mathfrak{k}(\mathfrak{sl}(2,\,\R))=\mathfrak{so}(2)$, respectively, which are the level zero subalgebras
  of $\mathfrak{k}(\frake_{10})$.
Likewise, we define
\begin{align}
\gS_A &= t_A+t^A = 2t^A, & \gS^{ab}&= K^a{}_b + K^b{}_a
\end{align}
as basis elements of the coset $\frake_{10} \ominus
 \mathfrak{k}(\frake_{10})$ at level zero.
Note that there is no $\gJ_A$ or $\gS_{IJ}$;
\textit{the indices on $\gJ_{IJ}$ and $\gS_{A}$ should
not be considered as split $E_8$ indices, but as pure $SO(16)$
indices}. This means that we raise the vector indices
$I,\,J,\ldots$ with the invariant $SO(16)$ metric $\delta^{IJ}$, so that $\gJ_{IJ}=\gJ^{IJ}$. On the other hand, $t_{IJ}=-t^{IJ}$, since we consider $t^{IJ}$ as an $\frake_8$ element. (For the spinor indices $A,\,B,\,\ldots$, upstairs and downstairs does not matter.)

Leaving level zero, the basis elements of $\mathfrak{k}(\frake_{10})$ and the coset will mix between positive and negative levels so the graded structure (\ref{gradstrukt}) will not be preserved,
\begin{align}
\gS^a{}_{\vA}&=E^a{}_{\vA}+F_a{}^{\vA}, &
\gS&=E+F, \nn\\
\gS^{ab}{}_{\vA}&=E^{ab}{}_{\vA}+F_{ab}{}^{\vA} &
  \gS_{\vA\vB}&=E_{\vA\vB}+F^{\vA\vB}. \nn\\
%  \end{align}
%  \begin{align}
\gJ^a{}_{\vA}&=E^a{}_{\vA}-F_a{}^{\vA}, &
\gJ&=E-F, \nn\\
\gJ^{ab}{}_{\vA}&=E^{ab}{}_{\vA}-F_{ab}{}^{\vA} &
  \gS_{\vA\vB}&=E_{\vA\vB}-F^{\vA\vB}.
  \end{align}
Computing the Cartan-Killing norm for these basis elements,
\begin{align} \label{forstaraden}
\la \gS_A | \gS_B \ra &=  4\delta_{AB},  &
\la \gJ_{IJ} | \gJ_{KL} \ra  &=-8\delta_{IK}\delta_{JL}, %2(\eta_{\vA\vB}+ \delta^{\vA}{}_{\vB}),
\nn\\
\la \gS^{ab} | \gS^{cd} \ra & = 4(\delta^{ac}\delta^{bd}-\delta^{ab}\delta^{cd}), &
\la \gJ^{ab} | \gJ^{cd} \ra &=
-4\delta^{ac}\delta^{bd},\nn\\
%\end{align}
%\begin{align*}
\la \gS^{a}{}_\vA | \gS^{b}{}_\vB \ra &=-\la \gJ^{a}{}_\vA | \gJ^{b}{}_\vB \ra= 2 \delta^{ab}\delta_{\vA}{}^{\vB},&
\la \gS_{\vA\vB} | \gS_{\vC\vD} \ra &=-\la \gJ_{\vA\vB} | \gJ_{\vC\vD} \ra
=28\mathbb{P}_{\vA\vB}{}^{\vC\vD},\nn\\
%\end{align*}
%\begin{align}
\la \gS^{ab}{}_{\vA} | \gS^{cd}{}_{\vB} \ra &=
-\la \gJ^{ab}{}_{\vA} | \gJ^{cd}{}_{\vB} \ra =
2\delta^{ac}\delta^{bd}\delta_{\vA}{}^{\vB}, &
\la \gS | \gS \ra &= -\la \gJ | \gJ \ra={2},
\end{align}
we see that the subspace $\mathfrak{k}(\mathfrak{e}_{10})$ is
negative-definite and that
$\mathfrak{e}_{10} \ominus \mathfrak{k}(\mathfrak{e}_{10})$
is positive-definite away from level zero.
Although some of the equations above are written in $E_8$ indices, for convenience, the position of the indices shows that they are in fact not
$E_8$ covariant.
The $E_8$ indices must be split into $SO(16)$ indices in order to give covariant equations.

\subsection{The non-linear sigma model}
Following \cite{Damour:2002cu, Damour:2004zy}
we now introduce a one-dimensional non-linear sigma-model based on the coset
$E_{10}/K(E_{10})$. The fields are represented by an $E_{10}$ valued
group element ${\cal V}(t)$, depending on a parameter $t$. This group element is subject to global $E_{10}$
transformations from the left and to the local subgroup $K(E_{10})$
from the right:
 \bea\label{nonlin}
  {\cal V}\longrightarrow g\hspace{0.2em}{\cal V}\hspace{0.2em}h(t)\;, \qquad \qquad g\in E_{10}\;, \quad
  h(t)\in K(E_{10})\;.
 \eea
Consequently, the $E_{10}$ invariant Maurer-Cartan forms are given
by ${\cal V}^{-1}\partial_t{\cal V}$. These can be decomposed into
compact and non-compact parts,
 \bea
  {\cal V}^{-1}\partial_t{\cal V} \ = \ {\cal P}(t) + {\cal Q}(t)\;,
  \qquad
  {\cal P}\in\frak{e}_{10}\ominus\frak{k}(\frak{e}_{10})\;, \qquad {\cal
  Q}\in\frak{k}(\frak{e}_{10})\;.
 \eea
While ${\cal P}$ and ${\cal Q}$ are $E_{10}$ invariant, they
transform under an infinitesimal local transformation $\delta{\cal
V}={\cal V}\hat{h}$, where $\hat{h}\in\frak{k}(\frak{e}_{10})$, as
 \bea
  \delta{\cal Q} \ = \ \partial_t\hat{h}+[{\cal Q},\hat{h}]\;,
  \qquad
  \delta{\cal P} \ = \ [{\cal P},\hat{h}]\;,
 \eea
i.e.~${\cal Q}$ is a (composite) gauge connection, while ${\cal P}$
transforms covariantly. The invariant action is then given by
 \bea\label{1dsigma}
  S \ = \ \frac14\int dt\, n(t)^{-1}\la{\cal
  P}(t)|{\cal P}(t)\ra \;,
 \eea
where 
$\la\ |\ \ra$ denotes the Cartan-Killing
form on ${\mathfrak{e}}_{10}$. Here, $n(t)$ is the lapse function establishing invariance
under the one-dimensional diffeomorphisms
 \bea
  \delta_{\xi}n \ = \ \xi\partial_t n + (\partial_t\xi) n\;, \qquad
  \delta_{\xi}{\cal P} \ = \ \xi\partial_t{\cal P}+ (\partial_t\xi) {\cal P}\;.
 \eea
The equations of motion obtained from (\ref{1dsigma}) are
 \bea\label{sigmabwgl}
  n\partial_t(n^{-1}{\cal P}(t))+[{\cal Q}(t),{\cal P}(t)] \ = \
  0\;,
 \eea
and the Hamiltonian constraint
 \bea
  \la\cP(t)|\cP(t) \ra \ = \ 0\;,
 \eea
which imply together that the motion follows a null geodesic.

So far our discussion was rather general. 
We are now going
to evaluate (\ref{1dsigma}) for the case we are interested in,
namely maximal supergravity in $D=3$. For this we use the level decomposition of $\mathfrak{e}_{10}$ with respect to $\sl(2,\,\R)\oplus \mathfrak{e}_8$ that we described in the preceding section.

The local $K(E_{10})$ invariance allows us to choose a suitable gauge for the
$E_{10}$-valued group element $\vV$. In the \textit{Borel gauge}, we can write
$\vV$ as a product
\begin{eqnarray}\label{representative}
\vV=\vV_{\ell}\vV_0=e^X(e^he^\vH),
\end{eqnarray}
where $\vV_{\ell}$ and $\vV_0$ are group elements corresponding to $\ell>0$ and
$\ell=0$, respectively. Thus we can expand the corresponding algebra elements in the basis of $\mathfrak{e}_{10}$ as
\bea
\label{cosel}
&&X=A_m{}^{\vM}E^m{}_{\vM}+B_{mn}{}^{\vM}E^{mn}{}_{\vM}+BE+B^{\vM\vN}E_{\vM\vN}
+\cdots, \\
&& h=h_a{}^bK^a{}_b\,,\quad
\vH=\vH_{\vA}t^{\vA}\,.
\eea
Here and in the following, 
$m,n,\ldots=1,\,2$ and $\cM,\,\cN \ldots = 1,\,2,\,\ldots,\,248$ denote
curved $GL(2)$ and $E_8$ indices, respectively.
This means that they are `world' indices indicating rigid transformations from
the left, while $\cA$ and $a$ are flat indices. 

In (\ref{representative}), the ordering of the exponentials is fixed
by the requirement that the fields $A_{m}{}^{\cM}$, etc.~transform
under the $SL(2,\,\R)$ according to their world indices
$m,\,n$.
In fact, under (\ref{nonlin}) we
have
 \bea
  {\cal V}_0\rightarrow g{\cal V}_0 h(t)\;, \qquad
  {\cal V}_{\ell}\rightarrow g{\cal V}_{\ell} g^{-1}\;.
 \eea
Therefore, parameterizing $g=\exp(R_{m}{}^nK^m{}_n)$
and using
$gA^mg^{-1}=A^nR_n{}^m$,
one finds
 \bea
  A_m{}^{\prime} \ = \ R_m{}^n A_{n}\;, \quad {\rm etc.}\;,
 \eea
as required (where we have omitted the $E_8$ indices).
In the Borel gauge, $\vP$ and $\vQ$ have the same components in the bases of
$\mathfrak{k}(\mathfrak{e}_{10})$ and the coset, except at level zero, 
\begin{align} \label{qpijs}
\vP&=P^{A}\gS_{A}+\tfrac12 P_{ab}\gS^{ab}
+P_a{}^{\vA}\gS^a{}_{\vA}+P_{ab}{}^{\vA}\gS^{ab}{}_{\vA}
+P\gS+P^{\vA\vB}\gS_{\vA\vB},\nn\\ 
\vQ&=\tfrac12Q^{IJ}\gJ_{IJ}+\tfrac12Q_{ab}\gJ^{ab}
+P_a{}^{\vA}\gJ^a{}_{\vA}+P_{ab}{}^{\vA}\gJ^{ab}{}_{\vA}+P\gJ+P^{\vA\vB}\gJ_{\vA\vB}.
\end{align}

We write $\vV_0$
as a product of two `vielbeine' $\exp{h}$ and $\exp{\vH}$, which are group
elements 
of $\gl(2,\,\R)$ and
$\mathfrak{e}_{8}$, respectively.
We denote the components of these group elements by $e_m{}^a$ and
$\vE^\vM{}_\vA$. 
Occasionally, we will denote the components of the inverses by $e_a{}^m$ and $\vE^\vA{}_\vM$. (The position of flat and curved indices thus keeps this notation unambiguous.)
Now we can write the components of $\vP$ and $\vQ$, defined by (\ref{qpijs}),
at level zero as
\begin{align}
P_{ab}&=\tfrac{1}{2}(e_a{}^m\pa_t{e}{}_m{}^b+e_b{}^m\pa_t{e}{}_m{}^a),
&P^A&=\tfrac12(\vE^{-1} \pa_t \vE)^A,
\nn\\
Q_{ab}&=\tfrac{1}{2}(e_a{}^m\pa_t{e}{}_m{}^b-e_b{}^m\pa_t{e}{}_m{}^a),
&Q^{IJ}&=\tfrac14(\vE^{-1} \pa_t \vE)^{IJ},
\end{align}
and we obtain the level zero part of the Lagrangian,
\begin{align} \label{ungaugedD=1}
\mathcal{L}_0&=
n^{-1}P^A P^A + \tfrac14n^{-1}(P_{ab}P_{ab}-P_{aa}P_{bb}).
\end{align}
As we will see below, this precisely coincides with the truncation
of ungauged supergravity to a one-dimensional time-like system.

We now turn to the computation of the full Maurer-Cartan form, including also
the $\ell>0$ part.
We then have
\begin{align}
\vV^{-1}\pa_t\vV &=
\vV_0{}^{-1} \pa_t \vV_0+
\vV_0{}^{-1}(\vV_\ell{}^{-1} \pa_t \vV_\ell) \vV_0.%\nn\\
\end{align}
The first term is the $\ell =0$ contribution which we used above. To
evaluate the second term we make use of the Baker-Campbell-Hausdorff
formulas
 \bea\label{BCH}
 \begin{split}
  e^{-A}de^{A} \ &= \
  dA+\tfrac{1}{2!}[dA,A]+\tfrac{1}{3!}[[dA,A],A]]+\cdots\;, \\
  e^{-A}Be^{A} \ &= \ B+[B,A]+\tfrac{1}{2!}[[B,A],A]+\cdots\;,
 \end{split}
 \eea
and find
 \begin{eqnarray}\nonumber
  {\cal V}_0{}^{-1}({\cal
  V}_{\ell}{}^{-1}\partial_t{\cal V}_{\ell}){\cal V}_0 &=&
  e_a{}^m\vE^\vA{}_\cM
D_tA_m{}^{\cM}E^a{}_\vA+
e_a{}^m
e_b{}^n\vE^\vA{}_\vM D_tB_{mn}{}^{\cM}E^{ab}{}_\vC
\nn\\&&+(\det{e})^{-1}(D_tBE
+14
\vE^\vA{}_\vM \vE^\vB{}_\vN D_tB{}^{\cM\cN}E^{\vA\vB}). \label{detekv}
 \end{eqnarray}
The determinant of the vielbein $e_m{}^a$ appears since the level two fields
$B$ and $B^{\vM\vN}$ transform with a nonzero weight under $\gl(2,\,\R)$.\footnote{More explicitly, the expansion gives
  \bea
  \cV_0^{-1} (D_t B E) \cV_0 &=& D_t B \left( E - h^a{}_b [K^b{}_a, E] +\tfrac12 h^a{}_b h^c{}_d [K^b{}_a,[K^d{}_c,E]]+ \ldots \right) \nn\\
  &=& D_t B E\left( 1- h^a{}_a +\tfrac12 (h^a{}_a)^2 +\ldots \right) = (\det  e)^{-1} D_t B E\,.\nn
 \eea}
In (\ref{detekv}) we have introduced the `covariant derivatives'
\begin{align} \label{covder}
D_tA_{m}{}^{\vM} \ &= \ \partial_t A_{m}{}^{\vM},\nn\\
D_tB_{mn}{}^{\vP} \ &= \ \partial_t B_{mn}{}^{\vP}
+\tfrac{1}{2}f_{\vM\vN}{}^{\vP}A_{(m}{}^{\vM}\partial_t A_{n)}{}^{\vN},\nn\\
D_tB \ &= \ \partial_t B -\tfrac{1}{4}\ep^{ab}\eta_{\vM\vN}
A_{m}{}^{\vM}\partial_t A_{n}{}^{\vN},\nn\\
D_tB^{\vM\vN} \ &= \ \partial_t B^{\vM\vN}%X_{ab}{}_{\vC}
-\tfrac{1}{2}\ep^{mn}\mathbb{P}{}_{\vP\vQ}
{}^{\vM\vN}A_{m}{}^{\vP}\partial_t A_{n}{}^{\vQ}.
\end{align}
Note that the $\frak{e}_{10}$ algebra leads to non-trivial
Chern-Simons like terms inside the covariant derivatives.
For instance, acting with the group element
\begin{align}
g=\exp(\Lambda_m{}^{\cM}E^{m}{}_{\cM}+\Lambda^{\cM\cN}E_{\cM\cN}+\cdots)
\end{align}
on the coset representative (\ref{representative}) yields the
following global symmetry transformation on the fields
 \bea\label{globale10}
  \delta_{\Lambda}A_m{}^{\cM} \ = \ \Lambda_m{}^{\cM}\;, \qquad
  \delta_{\Lambda}B^{\cM\cN} \ = \ \Lambda^{\cM\cN}+
  \ft12\varepsilon^{mn}\Lambda_{m}{}^{\cP}A_{n}{}^{\cQ}\mathbb{P}_{\cP\cQ}{}^{\cM\cN}\;,
 \eea
which leaves (\ref{covder}) invariant.

In order to project onto the non-compact part ${\cal P}(t)$, we have
to
replace $x$ by
$\ft12(x+x^T)$.
Then
using (\ref{bilform}) and
inserting into (\ref{1dsigma})
yields the sigma model Lagrangian
\begin{align} \label{lagrange5}
\mathcal{L}&=\mathcal{L}_0 + \tfrac{1}{8}n^{-1}
(g^{mn} \vG_{\cM \cN}
D_tA_m{}^{\cM} D_tA_n{}^{\cN}
+g^{mp}g^{nq}\vG_{\cM \cN}
D_tB_{mn}{}^{\cM}D_tB_{pq}{}^{\cN})
\nn\\
&\quad\quad\,\,\,+
\tfrac18 n^{-1} (\det{g})^{-1}(D_tBD_tB+14
\vG_{\cM \cP}\vG_{\cN \cQ} D_tB{}^{\cM\cN} D_tB{}^{\cP\cQ}),
\end{align}
where $\vL_0$ now can be written as
\begin{align}
\mathcal{L}_0
&=
\tfrac{1}{960}n^{-1}\pa_t{\vG}^{\vM\vN}\pa_t{\vG}^{\vP\vQ}{\vG}_{\vM\vP}{\vG}_{\vN\vQ}
+
\tfrac{1}{16}n^{-1}\pa_t{g}_{mn}\pa_t{g}_{pq}({g}^{mp}{g}^{nq}-{g}^{mn}{g}^{pq})
\end{align}
and we introduced the (inverse) `metrics'
 \begin{align} \label{metrikekvationer}
 g^{mn}&=e_a{}^me_a{}^n, & \vG_{\cM\cN} \ &= \ %\delta_{\cA\cB}
 {\vE}^{\cA}{}_{\cM}{\vE}^{\cA}{}_{\cN}\;.
 \end{align}
We stress that for the `$E_{8}$ metric', the contraction is not performed 
by means of the $E_{8}$ invariant Cartan-Killing form, but instead with the 
ordinary delta symbol. Specifically, in the $SO(16)$ decomposition, this
`metric' (\ref{metrikekvationer}) and its inverse read
 \bea\label{so16}
\vG_{\cM\cN} \ &=& \ \ft12 {\vE}^{IJ}{}_{\cM}{\vE}^{IJ}{}_{\cN} 
  + {\vE}^{A}{}_{\cM}{\vE}^{A}{}_{\cN}\;, \nn\\
\vG^{\cM\cN} \ &=& \ \ft12 {\vE}^\cM{}_{IJ} {\vE}^\cN{}_{IJ} 
  + {\vE}^\cM{}_{A}{\vE}^\cN{}_{A}\;,
 \eea
whereas the contraction with the (indefinite) Cartan-Killing metric
(\ref{Cartan-Killing}) would give rise to a relative minus sign between
the two terms on the r.h.s., and simply reproduce the Cartan-Killing metric:
$\vE^\cM{}_\cA \vE^\cN{}_\cB \, \eta^{\cA\cB} = \eta^{\cM\cN}$.
The equation (\ref{so16}) is consistent with the local $SO(16)$ 
symmetry, in accordance with the contraction over \textit{flat} indices.
Likewise, the first equation in (\ref{metrikekvationer}) is 
consistent with the local $SO(2)$ symmetry.

We compare (\ref{lagrange5}) with the expression for the Lagrangian that we
get directly from (\ref{1dsigma}) and (\ref{qpijs}), 
\begin{align}
\mathcal{L} %\label{lagrange2}
&=\tfrac{1}{4}n^{-1} \la \vP | \vP \ra 
\label{lagrange1}
=\mathcal{L}_0
+ \tfrac{1}{2}n^{-1}
(P_a{}^{\vA}P_a{}^{\vA}
+P_{ab}{}^{\vA}P_{ab}{}^{\vA}%
+PP +14P^{\vA\vB}P^{\vA\vB}).
\end{align}
Here the contraction of $E_8$ indices is again made with the delta symbol, as
in 
(\ref{so16}).
Comparing
the expressions (\ref{lagrange1}) and (\ref{lagrange5}), we see
that the components of $\vP$ are the `covariant derivatives' in (\ref{covder})
converted to flat indices,
\begin{align}
P_a{}^\vA &= \tfrac12 e_a{}^m\vE^\vA{}_\vM D_tA_m{}^\vM,
& P_{ab}{}^\vA &= \tfrac12 e_a{}^m e_b{}^n \vE^\vA{}_\vM %D_t h
D_tB_{mn}{}^\vM,\nn\\
P&=\tfrac12 (\det{e})^{-1}D_tB, & P^{\vA\vB} &= \tfrac12 (\det{e})^{-1}\vE^\vA{}_\vM \vE^\vB{}_\vN
D_tB^{\vM\vN}.
\label{identification}
\end{align}

\subsection{Equations of motion}

We now work out the equations of motion that follow from the Lagrangian 
(\ref{sigmabwgl}). 
In the truncation to $|\ell| \leq 2$, they read{\allowdisplaybreaks{
\begin{subequations}
\begin{align}
\label{e10einst}
n{\partial_t}(n^{-1}P_{ab})&=-2P_{ac}Q_{bc}
-P_a{}^{IJ} P_b{}^{IJ}-2P_a{}^A P_b{}^A
-2P_{ac}{}^{IJ} P_{bc}{}^{IJ}
-4 P_{ac}{}^A P_{bc}{}^A\nn\\&\quad\,
+\de_{ab}\Big[P_c{}^{IJ} P_c{}^{IJ}
+2P_c{}^A P_c{}^A
+2P_{cd}{}^{IJ} P_{cd}{}^{IJ}
+4P_{cd}{}^A P_{cd}{}^A+2PP\nn\\
&\qquad\quad\quad+7( P^{IJ\,KL}P^{IJ\,KL}+4P^{A\,IJ}P^{A\,IJ}+4P^{AB}P^{AB})\Big],\\
\label{e10scalars}
n{\partial_t}(n^{-1}P^A)&=\tfrac12\Ga^{IJ}{}_{AB}
(P^B Q^{IJ}
+P_a{}^BP_a{}^{IJ}
+P_{ab}{}^BP_{ab}{}^{IJ}\nn\\&\quad\quad\quad\quad\,\,
+28P^{BC}P^{IJ\,C}+14P^{B\,KL}P^{IJ\,KL}),\\
\label{e10lev1A}
n\pa_t(n^{-1}P_a{}^A)&=(P_{ab}-Q_{ab})P_b{}^A+
\tfrac12\Ga^{IJ}{}_{AB}(Q^{IJ}P_a{}^B+P_a{}^{IJ} P^B)\nn\\
&\quad\, -\tfrac12\Ga^{IJ}{}_{AB}(P_{ab}{}^BP_b{}^{IJ}+ P_{ab}{}^{IJ}P_b{}^B)
\nn\\&\quad\,
-\ep_{ab}(28P^{AB}P_b{}^B+14P^{A\,IJ}P_b{}^{IJ}+PP_b{}^A),\\
\label{e10lev1IJ}
n\pa_t(n^{-1}P_a{}^{IJ})&=(P_{ab}-Q_{ab})P_b{}^{IJ}
-4Q^{IK}P_a{}^{JK}+\Ga^{IJ}{}_{AB}P_a{}^AP^B\nn\\
&\quad\,-4P_{ab}{}^{IK}P_b{}^{JK}-\Ga^{IJ}{}_{AB}P_{ab}{}^AP_b{}^B
\nn\\&\quad\,
-\ep_{ab}(28P^{IJ\,A}P_b{}^A+14P^{IJ\,KL}P_b{}^{KL}-PP_b{}^{IJ}),\\
n{\partial_t}(n^{-1}P_{ab}{}^{A})&=2(P_{ac}-Q_{ac})P_{cb}{}^{A}+
\tfrac12 Q^{IJ}\Ga^{IJ}{}_{AB}P_{ab}{}^B+\tfrac12P_{ab}{}^{IJ}\Ga^{IJ}{}_{AB}P^B,\\
n{\partial_t}(n^{-1}P_{ab}{}^{IJ})&= 2(P_{ac}-Q_{ac})P_{cb}{}^{IJ}
-4Q^{IK}P_{ab}{}^{JK}+\Ga^{IJ}{}_{AB}P_{ab}{}^A P^B,\\
n{\partial_t}(n^{-1}P^{AB})&=P_{aa}P^{AB}+Q^{IJ}\Ga^{IJ}{}_{AC}P^{BC}+P^{B\,IJ}
\Ga^{IJ}{}_{AC}P^C,\\
n{\partial_t}(n^{-1}P^{A\,IJ})&=P_{aa}P^{A\,IJ}+\tfrac12 Q^{KL}\Ga^{KL}{}_{AB}P^{B\,IJ}
-4Q^{KI}P^{A\,KJ}\nn\\
&\quad\,+\tfrac12P^{IJ\,KL}\Ga^{KL}{}_{AB}P^B+\Ga^{IJ}{}_{BC}P^CP^{AB},\\
n{\partial_t}(n^{-1}P^{IJ\,KL})&=P_{aa}P^{IJ\,KL}-4Q^{MK}P^{ML\,IJ}
-4Q^{MI}P^{MJ\,KL}\nn\\&\quad\,
+\Ga^{IJ}{}_{AB}P^{A\,KL}P^B
+\Ga^{KL}{}_{AB}P^{A\,IJ}P^B,\\
n{\partial_t}(n^{-1}P)&=P_{aa}P. \label{eomendeee}
\end{align}
\end{subequations}
In the above equations the irreducibility constraint (\ref{villkor}) on the level two
field $P^{\cA\cB}$ is not  
spelled out explicitly, but see (\ref{a-tensorer}) and (\ref{3875}) below.}}

The equations of motion can of course also be computed directly from the Lagrangian (\ref{lagrange5}), without using the commutation relations.
By varying the level two fields, we get
\begin{align} \label{highereom}
0&=\pa_t (n^{-1}g^{mp}g^{nq}\vG_{\cM \cN}D_tB_{pq}{}^{\cN}),\nn\\
0&=\pa_t (n^{-1}(\det{g})^{-1}D_tB),\nn\\
0&=\pa_t (n^{-1}(\det{g})^{-1}\vG_{\cM \cP}\vG_{\cN \cQ}D_tB{}^{\cP\cQ}),
\end{align}
and for the first level,
\begin{align} \label{forstanivan}
0&=\tfrac12 n^{-1}(g^{mp}g^{nq}\vG_{\vP\vQ}f_{\vM\vN}{}^\vP D_tA_n{}^\vN D_tB_{pq}{}^\vQ\nn\\&\quad\quad\qquad
-\tfrac12 \ep^{mn}\eta_{\vM\vN}(\det{g})^{-1}D_tA_n{}^\vN D_tB
\nn\\&\qquad\quad\quad
-14 \ep^{mn}(\det{g})^{-1} \vG_{\vM\vP}\vG_{\vN\vQ}D_tA_n{}^\vN D_tB^{\vP\vQ})
\nn\\&\quad\,
-\tfrac12 \pa_t\Big[n^{-1}(2g^{mn}\vG_{\vM\vN}D_tA_n{}^\vN 
-g^{mp}g^{nq}\vG_{\vP\vQ}f_{\vM\vN}{}^\vP A_n{}^\vN D_tB_{pq}{}^\vQ\nn\\&
\qquad\qquad\qquad
+\tfrac12\ep^{mn}\eta_{\vM\vN}(\det{g})^{-1}A_n{}^\vN D_tB
\nn\\&\qquad\qquad\qquad
+14 \ep^{mn}(\det{g})^{-1}\vG_{\vM\vP}\vG_{\vN\vQ}A_n{}^\vN D_tB^{\vP\vQ})\Big]\,.
\end{align}
We use the equations (\ref{highereom})
to rewrite the second half of (\ref{forstanivan}),
\begin{align} \label{forenklat2}
n \pa_t(n^{-1}g^{mn}\vG_{\vM\vN}D_tA_n{}^\vN) 
&=
g^{mp}g^{nq}\vG_{\vP\vQ}f_{\vM\vN}{}^\vP D_tA_n{}^\vN D_tB_{pq}{}^\vQ
\nn\\&
\quad\,
-\tfrac12\ep^{mn}\eta_{\vM\vN}(\det{g})^{-1}D_tA_n{}^\vN D_tB\nn\\&\quad\,
-14 \ep^{mn}(\det{g})^{-1}\vG_{\vM\vP}\vG_{\vN\vQ}D_tA_n{}^\vN D_tB^{\vP\vQ}.
\end{align}
It is then straightforward to show that we
get the same equations as above. The equations (\ref{highereom})
can also be used
to rewrite the first half of (\ref{forstanivan}),
as we will see in section \ref{tltgs}.

\section{Gauged supergravity in three dimensions}

In this section we review gauged three-dimensional supergravity in a formulation suitable for comparison with the $E_{10}$ analysis of the preceding section. The comparison will be carried out in the next section.

The bosonic sector of ungauged maximal supergravity in three dimensions 
contains $128$ propagating scalars transforming in the coset 
$E_{8}/(Spin(16)/{\mathbb Z}_2)$ and a vielbein $\tte_\mu{}^\al$ that 
carries no dynamical degrees of freedom 
\cite{1983uft..conf..215J,Marcus:1983hb}. The scalars can also be described 
by an (internal) vielbein which we denote by $\ttE^\cM{}_\cA$ (which was 
denoted $\cV^\cM{}_\cA$ in \cite{Nicolai:2001sv}).\footnote{Generally, we will
  use the `typewriter' font for supergravity variables in order to distinguish
  them from the corresponding $E_{10}$ quantities.}
The inverses
will be written as $\tte_\al{}^\mu$ and $\ttE^\cA{}_\cM$.
 The curved indices
are written as Greek indices
$\mu,\nu,\ldots =(t,m)$ and the flat indices are
$\al,\,\be,\ldots=0,\ 1,\ 2$. The $E_8$ indices follow the same conventions as
before. We ignore fermions throughout the paper.

\subsection{The Lagrangian}

The construction of gauged three-dimensional supergravity where a subgroup
$G_0$ of the global symmetry group $E_{8}$ has been gauged proceeds via the
introduction of gauge fields $\ttA_\mu{}^\cM$ in the adjoint of $\frake_{8}$
such that one has the modified Maurer-Cartan
forms~\cite{Nicolai:2000sc,Nicolai:2001sv}\footnote{We reiterate that we have
  changed the normalization of the generators of the coset generators
  $\gS^A=2Y^A$, $\gJ^{IJ}=-2X^{IJ}$ compared to the generators used in
  \cite{Nicolai:2000sc,Nicolai:2001sv}. Also the space-time signature here is
  $(-++)$, opposite to that used there. The convention for the Levi-Civita
  symbol is $\ep^{012}=+1$.}
 \bea\label{sugracm}
  {\ttE}^{-1}\gD_{\mu}{\ttE} \ = \ttQ_\mu+\ttP_\mu =\ft12
  \ttQ_{\mu}^{IJ}\gJ^{IJ}+\ttP_{\mu}^{A}\gS^{A}\;,
 \eea
where the gauge-covariant derivative is given by
 \bea
  {\ttE}^{-1}\gD_{\mu}{\ttE} \ = \ {\ttE}^{-1}\partial_{\mu}{\ttE} + g\ttA_{\mu}{}^{\cM}\Theta_{\cM\cN}({\ttE}^{-1} t^{\cN} {\ttE})\;.
 \eea
The quantity $\Theta_{\cM\cN}$ is the constant embedding tensor
describing the generators of the Lie algebra $\frak{g}_0\subset
\frak{e}_{8}$ in terms of $\frak{e}_{8}$ generators: $X_\cM =
\Theta_{\cM\cN}t^\cN$.  There are only $\dim(\frak{g}_0)$ many
non-vanishing $X_\cM$ but it is convenient to maintain an
$E_{8}$ covariant notation. In such a notation, the embedding
tensor is symmetric in its indices and transforms in the ${\bf 3875
\oplus 1}$ representation of $E_{8}$. We will sometimes split it
into its irreducible parts as
 \bea
  \Theta_{\cM\cN} \ = \ \tilde{\Theta}_{\cM\cN}+\theta\,\eta_{\cM\cN}\;,
 \eea
where $\tilde{\Theta}$ transform in the ${\bf 3875}$, and $\theta$
is the singlet part.

Under infinitesimal local $G_0$ transformations with parameter $\Lambda^\cM X_\cM$ one has
 \begin{eqnarray}\label{gaugerig}
  \delta\, \ttA_{\mu}{}^{\cM} &=& \gD_{\mu}\Lambda^{\cM} \ \equiv \ \partial_{\mu}\Lambda^{\cM} +
  gf^{\cM\cN}{}_{\cK}\Theta_{\cN\cL}A_{\mu}{}^{\cL}\Lambda^{\cK}\;,\\
  \delta\, {\ttE} &=& g \Lambda^\cM X_\cM \,{\ttE}\;, \qquad
 \end{eqnarray}
and the Maurer-Cartan form is invariant. 

The bosonic Lagrangian of three-dimensional maximal gauged supergravity 
is~\cite{Nicolai:2000sc,Nicolai:2001sv}
\begin{align}\label{gaugedsugra}
\vL = \tte\left(\tfrac14 R- \ttP_{\mu}{}^A \ttP^{\mu}{}^A-V\right)
+ \vL_{CS},
\end{align}
with $\tte = \det(\tte_\mu{}^\alpha)$ and the Chern-Simons term
\begin{align}
\vL_{CS} = -\tfrac14 g\ep^{\mu\nu\rho} \Th_{\vM\vN}
\ttA_\mu{}^\vM\pa_\nu \ttA_\rho^\vN
-\tfrac{1}{12} g^2\ep^{\mu\nu\rho} \Th_{\vM\vN}
\Th_{\vP\vQ}
f^{\vM\vP}{}_{\vR} \ttA_{\mu}{}^\vN \ttA_{\nu}{}^\vQ \ttA_{\rho}{}^\vR\,.
\end{align}
Since there is no kinetic term for them, the gauge fields $\ttA_\mu{}^{\cM}$ 
do not contain propagating degrees of freedom. The gauging also introduces 
an indefinite scalar potential. In order to write it out, one introduces
the so-called T-tensor that transforms in the ${\bf 3875}$ of $E_8$, 
and is defined by 
\begin{align}
\tilde{T}_{\vA\vB}=\ttE^\vM{}_\vA \ttE^\vN{}_\vB \tilde{\Theta}{}_{\vM\vN}\,.
\end{align}
The field dependent T-tensor is thus the $E_8$ rotated version of the 
(constant) embedding tensor $\tilde{\Theta}_{\cM\cN}$. 
Note that here we have defined the T-tensor only with respect to ${\bf 3875}$, 
in contrast to \cite{Nicolai:2001sv}.   
The fact that $\tilde{T}$ transforms 
in the ${\bf 3875}$ implies that it has the components
\begin{align} \label{a-tensorer}
A_1{}^{IJ}&=-\de_{IJ}\theta+\tfrac17\tilde{T}_{IK\,JK},\nn\\
A_2{}^{I\dot{A}}&=-\tfrac17\Ga^{J}{}_{A\dot{A}}\tilde{T}_{IJ\,A},\nn\\
A_3{}^{\dot{A}\dot{B}}&=2\de_{\dot{A}\dot{B}}\theta+\tfrac1{48}
\Ga^{IJKL}{}_{\dot{A}\dot{B}}\tilde{T}_{IJ\,KL},
\end{align}
corresponding to the decomposition
\begin{align}\label{3875}
{\bf 3875} \rightarrow {\bf 135} \oplus {\bf 1820} \oplus {\bf 1920}
\end{align}
of this $E_8$ representation under $SO(16)$ \cite{Nicolai:2001sv}.
Here $A_1{}^{IJ}$ is symmetric, 
$A_1{}^{IJ}=A_1{}^{(IJ)}$ and $A_2{}^{I\dot A}$ traceless, that is
$\Gamma{}^I{}_{A\dot A} A_2{}^{I\dot A} = 0$. The potential then is the sum of 
two parts \cite{Nicolai:2001sv}, one negative-definite and the other 
positive-definite,
\begin{align} \label{apot}
V=\tfrac18 g^2 (-A_1{}^{IJ}A_1{}^{IJ}+\tfrac12 A_2{}^{I\dot{A}}
A_2{}^{I\dot{A}}).
\end{align}
Note that there is no contribution involving $A_3^{\dot A \dot B}$.
Alternatively, the potential can be written in the form
\begin{align}\label{gpot}
V & = \tfrac{1}{32} g^2G^{\cM\cN,\cK\cL} \Theta_{\cM\cN}
\Theta_{\cK\cL} \;,
\end{align}
where~\cite{Bergshoeff:2008qd}
 \bea\label{potG}
  G^{\cM\cN,\cK\cL} \ = \ \tfrac{1}{14} \ttG^{\cM\cK} \ttG^{\cN\cL}
        + \ttG^{\cM\cK} \eta^{\cN\cL}
        - \tfrac{3}{14} \eta^{\cM\cK} \eta^{\cN\cL}
        -\tfrac{4}{6727}\eta^{\cM\cN}\eta^{\cK\cL} \;
        \eea
with the metric $\ttG^{\cM\cN}$ defined in (\ref{so16}), but here with respect
to the supergravity $E_8$ vielbein $\ttE^\cM{}_\cA$.
Inserting (\ref{a-tensorer}) into (\ref{apot}), and using the
relations (\ref{enklarevillkor}) (which follow from the fact that 
$\tilde{T}$ transform in the ${\bf 3875}$ representation) we get yet another
expression for the potential,
\begin{align} \label{tpot}
V=\tfrac1{112} g^2(3 \tilde{T}_{AB}\tilde{T}_{AB} 
+\tilde{T}_{A\,IJ}\tilde{T}_{A\,IJ}- \tilde{T}_{IJKL}\tilde{T}_{IJKL})
-2 g^2\theta^2.
\end{align}
Both (\ref{gpot}) and (\ref{tpot}) will be used for the comparison with the
$E_{10}$ sigma model. Note, however, that in this form the decomposition
(\ref{3875})  is only implicit.

\subsubsection{Equations of motion}

Varying (\ref{gaugedsugra}) with respect to the gauge field one obtains the following non-abelian duality relation
 \bea\label{covdual}
  \tte^{-1}\varepsilon^{\mu\nu\rho}\Theta_{\cM\cN}\ttF_{\nu\rho}{}^{\cN} \ = \
  -4\Theta_{\cM\cN}{\ttE}^{\cN}{}_A \ttP^{\mu A} \,,
 \eea
 in terms of the non-abelian field strength
 \begin{align}
\ttF_{\mu\nu}{}^\vM = \pa_\mu \ttA_{\nu}{}^\vM - \pa_\nu \ttA_{\mu}{}^\vM
+g\Th_{\vP\vQ}f^{\vM\vP}{}_{\vR}\ttA_{\mu}{}^\vQ \ttA_{\nu}{}^\vR\,.
\end{align}
We stress that the summation in (\ref{covdual}) is only over the coset indices
$A$ and not over the whole $E_8$.
The Einstein equation can be written as
\bea\label{einstein}
R_{\mu\nu} = 4\, \ttP_\mu{}^A \ttP_\nu{}^A + 4\,{\mathtt{g}}_{\mu\nu} V \,,
\eea
where, again, the summation only is over the
$SO(16)$ spinor indices.

For the scalars, we first consider only the positive term in (\ref{apot}), and its variation along the coset,
\begin{align}
\delta(A_2{}^{I\dot{A}}
A_2{}^{I\dot{A}})
=\tfrac1{14}\Ga^{IJ}{}_{AB}(2\tilde{T}_{AC}\tilde{T}_{IJ\,C}+\tilde{T}_{A\,KL}
\tilde{T}_{IJ\,KL})
(\ttE^{-1}\de\ttE)^B.
\end{align}
Since we also have
\begin{align}
\delta(\ttP_\mu{}^A \ttP^\mu{}^A)
=\ttP{}_\mu{}^A\pa^\mu(\ttE^{-1}\de\ttE)^A+\tfrac12 \ttQ_\mu{}^{IJ}\Ga{}^{IJ}{}_{AB}\ttP^{\mu}{}^B(\ttE^{-1}\de\ttE)^A,
\end{align}
it follows that the scalar equation of motion, without the contribution 
from the negative term in the potential, becomes
\begin{align} \label{sugrascalarshalva}
\tte^{-1}\pa_\mu (\tte \ttP^\mu{}^A)
=\tfrac12\Ga^{IJ}{}_{AB}(\ttQ_\mu{}^{IJ}\ttP^\mu{}^B
-\tfrac1{56}g^2\tilde{T}_{BC}\tilde{T}_{IJ\,C}-\tfrac1{112}g^2\tilde{T}_{B\,KL}
\tilde{T}_{IJ\,KL})
+ \dots
\end{align}
For the negative-definite part in (\ref{apot}) we have
\begin{align}
\delta(A_1{}^{IJ}
A_1{}^{IJ})
=\tfrac1{14}\Ga^{IJ}{}_{AB}(-3\tilde{T}_{AC}\tilde{T}_{IJ\,C}+2\tilde{T}_{A\,KL}
\tilde{T}_{IJ\,KL})
(\ttE^{-1}\de\ttE)^B.
\end{align}
Thus the full 
equation of motion for the scalars reads
\begin{align}\label{sugrascalars}
\tte^{-1}\pa_\mu (\tte \ttP^\mu{}^A)
=\tfrac12\Ga^{IJ}{}_{AB}(\ttQ_\mu{}^{IJ}\ttP^\mu{}^B
+\tfrac1{14}g^2\tilde{T}_{BC}\tilde{T}_{IJ\,C}-\tfrac3{112}g^2\tilde{T}_{B\,KL}\tilde{T}_{IJ\,KL}).
\end{align}
This rewritten form of the equations of motion of
\cite{Nicolai:2000sc,Nicolai:2001sv} is convenient for the comparison
with the $E_{10}$ sigma model.

\subsubsection{Constraints}

From the form of the Maurer-Cartan form (\ref{sugracm}) one deduces the following integrability relations
\begin{align}
g \ttF_{\mu\nu}{}^\cM \Theta_{\cM\cN} \ttE^\cN{}_\cA t^\cA
  = 2 \partial_{[\mu} \ttP_{\nu]} + 2 \partial_{[\mu} \ttQ_{\nu]}
   + \left[\ttQ_\mu+\ttP_\mu,\ttQ_\nu+\ttP_\nu\right]   \,.
\end{align}
Using the duality relation (\ref{covdual}) this can be rewritten as a relation expressed solely in terms of $\ttP$, $\ttQ$ and the embedding tensor as
\begin{align}
\label{constraint1}
2\partial_{[\mu} \ttP_{\nu]} +  2\partial_{[\mu} \ttQ_{\nu]}
&=
-\left[\ttQ_\mu+\ttP_\mu,\ttQ_\nu+\ttP_\nu\right]
  \nn\\
&\quad\,
+ \tte g \varepsilon^{\rho}{}_{\mu\nu} \tilde{T}_{A\,IJ} \ttP_\rho{}^A t^{IJ}
+ 2\tte g \varepsilon^{\rho}{}_{\mu\nu} \tilde{T}_{AB} \ttP_\rho{}^A t^B
+ 2\tte g \varepsilon^{\rho}{}_{\mu\nu} \theta \ttP_\rho{}^A t^A\,.
\end{align}
The equation (\ref{constraint1}) is the deformation of the usual integrability constraint of non-linear sigma models in the presence of gauging.  In addition there are three-dimensional Bianchi constraints, {\em viz.}
\bea\label{constraint2}
\Theta_{\cM\cN}\gD_{[\mu} \ttF_{\nu\rho]}{}^\cN = 0 \,
\eea
for the gauge field and for the gravity sector
\bea\label{constraint3}
R_{[\mu\nu\, \rho]\sigma} = 0\,.
\eea

Finally, the embedding tensor is subject to linear and quadratic constraints~\cite{Nicolai:2000sc,Nicolai:2001sv}. The linear constraint arises from supersymmetry and implies that it transforms in the ${\bf 1\oplus 3875}$ part of the symmetric tensor product of two ${\bf 248}$ representations, so that the ${\bf 27000}$ is absent.
This constraint leads to the relations (\ref{enklarevillkor}) that we already used in (\ref{tpot}) and (\ref{sugrascalars}) to simplify expressions involving the $T$ tensor.
The quadratic constraint reads
 \bea\label{quadconstr}
  \cQ_{\cM\cN,\cP} \ \equiv \ \Theta_{\cK\cP}\Theta_{\cL(\cM}f^{\cK\cL}{}_{\cN)} \ = \ 0\;.
 \eea
As we will see in section~\ref{redsec}, further constraints on the fields
arise when some of the gauge freedom has been fixed.

\subsubsection{Reformulation with deformation and top-form potentials}
Here we briefly introduce a reformulation of gauged supergravity
with so-called deformation and top-form
potentials~\cite{deWit:2008ta,Bergshoeff:2007vb}, which will be
useful for the interpretation of the $E_{10}$ equations below. These
potentials are part of a tensor hierarchy introduced in \cite{de
Wit:2005hv} and can be viewed as Lagrange multipliers enforcing the
constancy of the embedding tensor and the quadratic constraint.
Denoting the deformation two-form by $\ttB_{\mu\nu}{}^{\cM\cN}$ 
and the top-form by $\ttC_{\mu\nu\rho}{}^{\cM\cN,\cP}$, which respectively
transform in the $\bf{1}\oplus\bf{3875}$ and $\bf{3875}\oplus\bf{147250}$
representations of $E_8$ 
\cite{deWit:2008ta,Bergshoeff:2007vb}, one has
 \bea\label{defaction}
  {\cal L}_{\rm tot} \ = \ {\cal L}_g+
  \ft14 g\varepsilon^{\mu\nu\rho}\gD_{\mu}\Theta_{\cM\cN}
  \ttB_{\nu\rho}{}^{\cM\cN} -
  \ft16g^2\Theta_{\cK\cP}\Theta_{\cL(\cM}f^{\cK\cL}{}_{\cN)}
  \varepsilon^{\mu\nu\rho}\ttC_{\mu\nu\rho}{}^{\cM\cN,\cP}\;,
 \eea
where the embedding tensor now satisfies only the linear constraint.
Here we have written a covariant derivative on $\Theta_{\cM\cN}$,
 \bea
  \gD_{\mu}\Theta_{\cM\cN} \ = \ \partial_{\mu}\Theta_{\cM\cN}
  +2g\ttA_{\mu}{}^{\cP}\Theta_{\cK\cP}\Theta_{\cL(\cM}f^{\cK\cL}{}_{\cN)}\;.
 \eea
The second term vanishes identically upon use of the quadratic
constraint, whence the equations of motion imply that $\Theta$ is
constant (and not just covariantly constant).
Since the space-time dependent embedding tensor is now a
dynamical field, it possesses its own equations of motions, which
can be viewed as duality relations between the 2-form potential and
the embedding tensor \cite{deWit:2008ta,Bergshoeff:2008qd}. Below we
will see that an analogous relation follows naturally from the
sigma model equations of motion, with the $E_{10}$ field $B^{\cM\cN}$ 
interpreted as (the Hodge dual of) the spatial part of  the deformation
potential.  
By contrast, in the $E_{11}$ approach of \cite{Riccioni:2007ni} both
$B_{\mu\nu}{}^{\cM\cN}$ and $C_{\mu\nu\rho}{}^{\cM\cN,\cP}$ appear in
the decomposition of $E_{11}$, whereas the embedding tensor must 
be introduced as an `extraneous' object to parametrize the deformation of 
the derivative in the Cartan form.

\subsection{Dimensional reduction to $D=1$}
\label{redsec}

We now effectively reduce the three-dimensional gauged supergravity theory to a
one-dimensional time-like system. For this we perform the ADM-like
split of the vielbein 
 \bea\label{vielbein}
  \tte_{\mu}{}^\al \ = \ 
  \left(\begin{array}{cc} N &
  0 \\ 0 & \tte_{m}{}^a
  \end{array}\right)\;,
 \eea
%in $D=3$
in which everything depends only on one coordinate $x^0 =t$ and we
have split curved indices as $\mu=(t,m)$ and flat ones as
$\al=(0,a)$ (with signature $(-++)$). Here we have chosen a gauge
with vanishing shift $N^m$, which turns out to be necessary in
order to match the $E_{10}$ coset. As stressed before, gauge fixing is crucial for comparing the $E_{10}$ sigma model to supergravity. The field $\tte_{m}{}^{a}$
denotes the internal `spatial' vielbein, i.e.~an element of
$GL(2,\,\R)/SO(2)$. The three-dimensional Einstein-Hilbert Lagrangian in
(\ref{gaugedsugra}) can be rewritten up to a total derivative as 
 \bea
  \tfrac14 \tte R\ = \
  -\tfrac{1}{16}\tte\Omega^{\al\be\,\ga}\Omega_{\al\be\,\ga}
  +\tfrac{1}{8}\tte\Omega^{\al\be\,\ga}\Omega_{\be\ga\,\al} 
  +\tfrac{1}{4}\tte\Omega_{\al\be}{}^\be \Omega^{\al\ga}{}_\ga \;,
 \eea
where $\Omega_{\al\be\,\ga}$ are the coefficients of anholonomy:
 \bea
  \Omega_{\al\be\,\ga} =
  \tte_{\al}{}^{\mu}\tte_{\be}{}^{\nu}(\partial_{\mu}\tte_{\nu
  \ga}-\partial_{\nu}\tte_{\mu \ga})\;.
 \eea
The only non-vanishing components in the strict reduction to $D=1$ are
 \bea
  \Omega_{a  0\,b}=-\Omega_{0a\,b}=-N^{-1}\tte_{a}{}^m\partial_t
  \tte_{m b}=:-N^{-1}\tth_{ab}\;,
 \eea
where we have introduced the $\frak{gl}(2,\,\R)$--valued current
$\tth_{ab}$ converted into flat indices. The current has both a symmetric and an antisymmetric part, $\tth_{ab}=\ttP_{ab}+\ttQ_{ab}$.
Inserting into the Einstein-Hilbert action, one finds that the antisymmetric part cancels and the resulting expression is
 \bea
  \tte^{-1}{\cal L}_{\rm EH} \ = \ \ft14 N^{-2}
  \left(\ttP_{ab}\ttP_{ab}-\ttP_{aa}\ttP_{bb}
\right)\;.
 \eea
On the other hand, the $E_{8}$ valued fields are all scalars and
trivially reduce according to ${\ttE}(x)\rightarrow {\ttE}(t)$.
Using $\tte=\det (\tte_{\mu}{}^{\al})=N\det(\tte_m{}^a)$, one
finds in total for the case of ungauged supergravity
 \bea\label{reduction}
  {\cal L}^{D=1}_{g=0} \ = \ 
  \ttn^{-1}\ttP_t{}^A\ttP_t{}^A
  +\ft14 \ttn^{-1}\left(\ttP_{ab}\ttP_{ab}
  -\ttP_{aa}\ttP_{bb}\right) \;,
 \eea
where we have defined the quantity
\bea\label{lapse}
 \ttn=  N
(\det(\tte_m{}^a))^{-1}\,. \eea Evidently, (\ref{reduction}) has
exactly the same form as the level zero Lagrangian
(\ref{ungaugedD=1}).

We turn now to gauged supergravity. For the reduction of the tensor fields we choose a temporal gauge
\bea\label{tempgauge}
\ttA_t{}^\cM=0\;,\quad \ttB_{tm}{}^{\cM\cN} =0\;,\quad \ttC_{tmn}{}^{\cM\cN,\cP} = 0\,.
\eea
Reducing the action
(\ref{defaction}) of gauged supergravity to $D=1$, we then find
 \begin{eqnarray}\label{gaugered}
  {\cal L}_g^{D=1}  &=& {\cal L}_{g=0}^{D=1}
  -\ttn^{-1}N^2\ttg^{mn}[\ttE^{-1}\gD_m\ttE]^{A}[\ttE^{-1}\gD_n\ttE]^{A}
   - \ttn^{-1} N^2 V\\ \nonumber
   &&+\ft{1}4g\ep^{mn}\ttA_m{}^{\cM}\Theta_{\cM\cN}\partial_t
   \ttA_n{}^{\cN}
    +\tfrac{1}{4}g\varepsilon^{mn}\gD_t\Theta_{\cM\cN} \ttB_{mn}{}^{\cM\cN}\;.
 \end{eqnarray}
Here, $\gD_m\ttE$ denotes the spatial part of the gauge-covariant
derivative, which in the case of pure time dependence reads
 \bea
  \ttE^{-1}\gD_m\ttE \ = \
  g\ttA_m{}^{\cM}\Theta_{\cM\cN}\ttE^{-1}t^{\cN}\ttE\;.
 \eea
The appearance of the gauge vector here is the only remnant of the
gauging in the scalar kinetic terms. In fact, the gauge choices
(\ref{tempgauge}) have the advantage that the time component of the
gauge covariant derivatives in $D=1$ collapses, e.g. \bea
\ttE^{-1}\gD_t \ttE  \ \equiv \ \ttE^{-1}\partial_t \ttE\,.
\eea 
Similarly, the cubic term in the reduction of the Chern-Simons
term disappears as well as the top-form potential term enforcing the
quadratic constraint. That the Maurer-Cartan forms are unchanged is
essential for the comparison with the $E_{10}$ model in its original
form.

When fixing gauges one should not forget the equations of motion
(constraints) resulting from varying with respect to the temporal
components of the gauge fields in (\ref{tempgauge}). They read from
(\ref{covdual}) and (\ref{defaction})
 \bea
\label{gauss}  C_{\cM}  &:=&  \ttn^{-1} \varepsilon^{mn} \Theta_{\cM\cN} \ttF_{mn}{}^\cN + 4 \Theta_{\cM\cN} \ttE^\cN{}_A \ttP_t{}^A
    \ = \  0\;,\\
  \label{spattheta}
  C_{\cM\cN}^m &:=& \ttn^{-1}g\varepsilon^{mn} D_n \Theta_{\cM\cN} \ =\ 0\;,\\
  \label{quadtheta}
  C_{\cM\cN,\cP} &:=& g^2 \Theta_{\cK\cP}\Theta_{\cL(\cM}f^{\cK\cL}{}_{\cN)} \ = \ 0\;.
 \eea
 As constructed, the constraints for $\ttB_{tm}{}^{\cM\cN}$ and
 $\ttC_{tmn}{}^{\cM\cN,\cP}$ correspond to the (spatial) constancy of the
 embedding tensor and the quadratic constraint. Below we will interpret the
 temporal constancy of $\Theta_{\cM\cN}$ as an equation of motion rather than as
 a constraint.

\subsection{Beyond dimensional reduction}

The $E_{10}$ model also takes into account terms that are beyond dimensional
reduction to $D=1$~\cite{Damour:2002cu,Damour:2004zy}. Therefore we also need
to keep track of terms that arise from spatial gradients and contribute
to the equations of motion.
Instead of writing out all the resulting equations we illustrate the procedure
in the example of equation (\ref{constraint1}). Considering the equation in
flat spatial indices and split into $\frak{so}(16)$ and coset components we
find for the $(\al,\be)=(0,a)$ component 
\begin{align}
\pa_0 \ttQ_a{}^{IJ} - \pa_a \ttQ_0{}^{IJ}&=
-4\ttQ_0{}^{[I|K}\ttQ_a^{J]K}-
\Ga^{IJ}{}_{AB}\ttP_a{}^A\ttP_0{}^B\nn\\&\quad\,
-N^{-1}(\ttQ_{ab}+\ttP_{ab})\ttQ_b{}^{IJ}
 \label{sugralev1IJ}
- \tte g \ep_{ab}\tilde{T}^{A\,IJ}\ttP_b{}^A,\\
\label{sugralev1A}
\pa_0 \ttP_a{}^{A} - \pa_a\ttP_0{}^A &=
\tfrac12\ttQ_0{}^{IJ}\Ga^{IJ}{}_{AB}\ttP_a{}^B
-\tfrac{1}{2}\ttQ_a{}^{IJ}\Ga^{IJ}{}_{AB}\ttP_0{}^B\nn\\&\quad\,
-N^{-1}(\ttQ_{ab}+\ttP_{ab})\ttP_b{}^{A}
+ \tte g \ep_{ab}(\tilde{T}^{AB}+\de^{AB}\theta)\ttP_b{}^B\,.
\end{align}
In analogy with these equations spatial dependence can be retained
systematically in all equations.

\section{The supergravity/$E_{10}$ correspondence} \label{comparison}

In this section we compare (a certain truncation) of supergravity to
the $E_{10}$ coset model. First, as a consistency check, we compare
the dynamics of ungauged supergravity with only time dependence to
the $\ell =0$ truncation of the $E_{10}$ equations of motion. Then,
in section~4.2, we discuss ungauged supergravity with the inclusion of
certain spatial gradients, that should be related to the $\ell=1$
truncation of the $E_{10}$ theory. An alternative interpretation of
the $\ell=1$ state is as a gauge vector and so we discuss a possible
relation between gauged supergravity and $E_{10}$ in section~4.3. 
Finally, we analyze the possible
$E_{10}$ interpretation of the gauge constraints and quadratic
constraints on the supergravity side in section~4.4.

\subsection{Ungauged supergravity in $D=1$}

The equations of motion of ungauged supergravity reduced to only
time dependence follow from the Lagrangian displayed in
(\ref{reduction}). As this Lagrangian is identical to the $\ell=0$
part of the Lagrangian of the $E_{10}$ sigma model derived in
(\ref{ungaugedD=1}) and depends on the same fields, the associated
dynamics agrees trivially. The `dictionary' which achieves this
correspondence at level $\ell=0$ reads 
\bea\label{dict0} 
n(t) \equiv
\ttn(t)\,, &&  P_{ab}(t) \equiv \ttP_{ab}(t)\,,\quad
Q_{ab}(t) \equiv \ttQ_{ab}(t)\,,\nn\\
&&P^A(t) \equiv \ttP_t{}^A(t)\,,\quad Q^{IJ}(t) \equiv
\ttQ_t{}^{IJ}(t)\,, 
\eea 
where $\ttn(t)$ is defined in
(\ref{lapse}). Here, we have displayed the coset quantities on the
left hand side and the supergravity variables on the right hand side
-- one can also write the correspondence in terms of the coset
elements as \bea e_m{}^a(t) \equiv \tte_m{}^a(t)\,,\quad
\cE^\cM{}_\cA (t) \equiv \ttE^\cM{}_\cA(t)\,. \eea The only equation
besides the equations of motion here is the Hamiltonian constraint
and it is mapped to the null condition of the geodesic.

When relaxing the strict dimensional reduction we will retain this dictionary except that we will interpret the supergravity variables to be the values at a fixed spatial point ${\bf x}_0$, so that the dictionary modifies to
\bea\label{dict01}
n(t) \equiv \ttn(t,\,{\bf{x}}_0)\,, &&  P_{ab}(t) \equiv \ttP_{ab}(t,\,{\bf{x}}_0)\,,\quad
Q_{ab}(t) \equiv \ttQ_{ab}(t,\,{\bf{x}}_0)\,,\nn\\
&&P^A(t) \equiv \ttP_t{}^A(t,\,{\bf{x}}_0)\,,\quad Q^{IJ}(t) \equiv \ttQ_t{}^{IJ}(t,\,{\bf{x}}_0)\,,
\eea
or, in terms of the coset variables,
\bea
e_m{}^a(t) \equiv \tte_m{}^a(t,\,{\bf{x}}_0)\,,\quad \cE^\cM{}_\cA (t) \equiv \ttE^\cM{}_\cA(t,\,{\bf{x}}_0)\,.
\eea

\subsection{Level $\ell=1$ as spatial gradient}
Let us now turn on the fields at level $\ell=1$ of the coset model.
One possible interpretation here is that this corresponds to a
spatial gradient --- in contrast to the interpretation as a gauge
vector, which we will discuss in the next section. 
For the investigation of spatial gradients it turns out to be useful to compare
both sides of 
the correspondence not at the level of the elementary fields but instead at
the level of the derived object $\cP$ that carries flat indices.
By studying the
Einstein equation (\ref{e10einst}) and the equations of level
$\ell=1$, (\ref{e10lev1A}) and (\ref{e10lev1IJ}), one finds after
comparison with (\ref{einstein}), (\ref{sugralev1IJ}) and
(\ref{sugralev1A}) that the dictionary on this level is
\begin{align}
\label{dict1} P_a{}^A(t) &\equiv  N \ep_{ab}\, \ttP_b{}^A
(t,{\bf x}_0)\,,&%\quad 
P_a{}^{IJ}(t) &\equiv  -N \ep_{ab}\,
\ttQ_b{}^{IJ} (t,{\bf x}_0)\,. 
\end{align}

This choice together with
(\ref{dict0}) makes the sigma model equations match largely with the
supergravity equations in the absence of gauging, where now the
equations of motion at $\ell=1$ correspond to the integrability
constraints (\ref{sugralev1IJ}) and (\ref{sugralev1A}) of the
three-dimensional theory. There are, however, 
terms that do not quite match. First of all, the equation of motion
(\ref{e10einst}) gets translated into 
\bea \label{ell1einst}
R_{ab} = 2\,\ttP_a{}^A
\ttP_b{}^A +\,\ttQ_a{}^{IJ}\ttQ_b{}^{IJ} 
\eea 
if spatial gradients
of the spin connection are truncated as usual in such
correspondences~\cite{Damour:2004zy}. This is not the correct
Einstein equation, see (\ref{einstein}), in that the coefficient of
$\ttP_a{}^A\ttP_b{}^A$ is $2$ rather than $4$ and that there is an
extra term proportional to $\ttQ^2$. The first problem is
immediately related to a similar discrepancy in the $D=11$
interpretation of the $E_{10}$ model~\cite{Damour:2004zy} where one
contribution to the only spatial derivatives in the curvature term
in $D=11$ was missing.\footnote{More precisely, the spatial Ricci
tensor $R_{ab}$ in
  $D=11$ has contributions (eq.~(4.81) in~\cite{Damour:2004zy}) of the form
  \bea
  \frac14 \Omega_{cd\,a}\Omega_{cd\,b} -\frac12\Omega_{ac\,d}\Omega_{bc\,d}
  -\frac12\Omega_{ac\,d}\Omega_{bd\,c}
  \eea
  and it is the last term which is not reproduced by the
  sigma model. But it contributes to the scalar energy-momentum tensor in
  lower dimensions.}
 After reduction to $D=3$ this problem gets shifted into the
scalar sector which explains why the scalar energy-momentum tensor does not
have the right coefficient. The $\ttQ^2$ term arises in a similar way in the
sigma model and has no counterpart in supergravity (where it would violate
the invariance under local $SO(16)$). The same term was already
noticed in \cite{Kleinschmidt:2005bq}.

It is noteworthy that there are no difficulties with the spatial curvature in $D=3$ since the problematic term vanishes completely due to our gauge choice. Indeed, one has that the full spatial anholonomy is given by
\bea
\Omega_{ab\,c} = -\ep_{ab}\ep_{cd} \Omega_{de\,e}\,.
\eea
Since we always choose the trace $\Omega_{de\,e}$ to vanish, the full spatial anholonomy vanishes in $D=3$ and gives no contribution to the $\Omega^2$ terms in $R_{ab}$. In other words, in this gauge choice there is no dual graviton in agreement with its absence in the table
of representations of $E_{10}$ under $SL(2,\,\R)\times E_8$ (table \ref{reptable}).\footnote{Since gravity in $D=3$ is not propagating one would not have expected a dual graviton.}

The final equation of motion to be compared is the equation of motion for the scalars, (\ref{e10scalars}) on the $E_{10}$ side and (\ref{sugrascalars}) on the supergravity side. Here, we find agreement in the absence of gauging.

We would like to comment on the interpretation of the dictionary (\ref{dict1}). One can introduce dual vector fields to the $E_8$ coset scalars also in the absence of gauging, similar to the duality relation (\ref{covdual}). These vector fields are the ones that appear in coset element (\ref{cosel}) at level $\ell=1$.

\subsection{Level $\ell= 2$ and gauged supergravity} \label{tltgs}

In this section we turn to gauged supergravity. First, we employ the
interpretation that the level $\ell=1$ 
field is not related to (spatial derivatives of) scalars prior to
any gauging, but instead the genuine gauge field to be introduced on
top of the scalars. According to this picture
we will compare to a purely time-like truncation.
As the level $\ell=2$ fields naturally encode the gauging, they will
be used at the same time. In a second step we consider the inclusion of
spatial gradients in the presence of gauging. For this we will discuss
the extension of the dictionary 
(\ref{dict01}) and (\ref{dict1}) to level $\ell=2$.

We start from the gauged supergravity action (\ref{gaugered}),
reduced to one dimension. Since on the $E_{10}$ side there is no
analogue of the zero-component of the gauge field
$\ttA_{\mu}{}^{\cM}$, we use the gauge-fixing condition
$\ttA_{t}{}^{\cM}=0$. Moreover, it turns out to be convenient to
rewrite the action entirely in terms of the $E_{8}$ `metric'
$\ttG^{\cM\cN}$. For this we use the identity
 \bea
 {\ttE}^{\cM}{}_{A}{\ttE}^{\cN A} \ = \
  \ft12\left(\ttG^{\cM\cN}+\eta^{\cM\cN}\right)\;,
 \eea
which follows from the fact that the Cartan-Killing metric
$\eta^{\cM\cN}$ differs from $\ttG^{\cM\cN}$ by a relative sign in
the non-compact part. The Lagrangian (\ref{gaugered}) reads
 \begin{eqnarray}
  L_{g}^{D=1} &=& L_{g=0}^{D=1}-\ft18
  g^2\tte
  \ttg^{mn}(\ttG^{\cM\cN}+\eta^{\cM\cN})\Theta_{\cM\cK}\Theta_{\cN\cL}\ttA_{m}{}^{\cK}\ttA_n{}^{\cL}-\tte V \\ \nonumber 
  &&+\ft14g\Theta_{\cM\cN}\varepsilon^{mn}\ttA_{m}{}^{\cM}\partial_{t}\ttA_{n}{}^{\cN}\;.
 \end{eqnarray}
For convenience we have here used the conventional formulation
without deformation potential, as the field equations merely relate this
potential to the embedding tensor. In contrast, the analogous equations on the
$E_{10}$ side introduce the embedding tensor.

The `Einstein' equations obtained by varying with respect to the
spatial $\ttg^{mn}$ read
 \bea\label{Einsteineom}
  \begin{split}
   \frac{\delta L_{0}}{\delta
   \ttg^{mn}}&+\ft12 \tte \ttg_{mn}V\\
   &+\tfrac{1}{16}g^{2}\tte(\ttG^{\cM\cN}+\eta^{\cM\cN})\Theta_{\cM\cK}
   \Theta_{\cN\cL}\left(\ttg_{mn}\ttg^{kl}\ttA_{k}{}^{\cK}\ttA_{l}{}^{\cL}-2\ttA_{m}{}^{\cK}\ttA_{n}{}^{\cL}\right)
   \ = \ 0\;,
  \end{split}
 \eea
while for the scalar equations we find
 \bea\label{scalareom}
  \begin{split}
   \frac{\delta L_0}{\delta \ttG^{\cM\cN}}&-\ft18
   g^2 \tte\ttg^{mn}\Theta_{\cM\cK}\Theta_{\cN\cL}\ttA_{m}{}^{\cK}\ttA_{n}{}^{\cL}\\
   &-\tfrac{1}{7\cdot
   32}\tte
 g^{2}\ttG^{\cK\cL}\Theta_{\cM\cK}\Theta_{\cN\cL}
  -\tfrac{1}{16} \tte g^{2}\eta^{\cK\cL}\Theta_{\cM\cK}\Theta_{\cN\cL} 
   \ = \ 0\;,
  \end{split}
 \eea
using the explicit form of the scalar potential in
(\ref{potG}). Here we do not write out the variation of $L_0$, since
we verified already that this Lagrangian coincides on both sides of
the correspondence. Finally, varying with respect to the
non-propagating vector fields $\ttA_{m}{}^{\cM}$ yields the
one-dimensional form of the duality relation,
 \bea\label{vectordual}
   g\Theta_{\cM\cN}\varepsilon^{mn}\partial_{t}\ttA_{n}{}^{\cN}
   +\ft12g^{2}\tte
   (\ttG^{\cK\cL}+\eta^{\cK\cL})\ttg^{mn}\Theta_{\cM\cK}\Theta_{\cN\cL} 
   \ttA_{n}{}^{\cN} \ = \ 0\;.
 \eea

At first sight these equations are rather different from the
 sigma model equations, which are given by
 \bea\label{e10zero}
 \begin{split}
  \frac{\delta L_0}{\delta g^{mn}}&+\ft18 n^{-1}
  \vG_{\cM\cN}\partial_t A_m{}^{\cM}\partial_t
  A_n{}^{\cN}\\
  &+\ft18 n^{-1}(\det{g})^{-1}g_{mn}\left(D_tBD_tB+14\vG_{\cM\cP}
  \vG_{\cN\cQ}D_tB^{\cM\cN}D_tB^{\cP\cQ}\right) \ = \ 0\;,\\
  \frac{\delta L_0}{\delta G_{\cM\cN}}&+\ft18
  n^{-1}g^{mn}\partial_tA_m{}^{\cM}\partial_tA_{n}{}^{\cN}
  +\tfrac{14}{4}n^{-1}(\det{g})^{-1}\vG_{\cK\cL}D_tB^{\cM\cK}D_tB^{\cN\cL} \ = \ 0\;
 \end{split}
 \eea
for the $\ell=0$ fields, and by (\ref{highereom}) and
(\ref{forenklat2}) for the higher-level fields. Consistent with the
field equations, we set in the following $D_{t}B_{mn}{}^{\cM}=0$,
since their meaning will be discussed below.

We will see that the equations on both sides are more closely
related, if one uses the observation that in $D=1$ second-order
equations can be integrated to first-order equations. 
For instance,
the equation (\ref{highereom}) gives rise to integration constants which
can be identified with the components of the embedding tensor,
 \bea\label{embedd}
 \begin{split}
  n^{-1}(\det{g})^{-1}D_tB \ &= \
  c_1g\theta\;, \\
  n^{-1}(\det{g})^{-1}\vG_{\cM\cP}\vG_{\cN\cQ}D_tB^{\cP\cQ} \ &= \
  c_2g\tilde{\Theta}{}_{\cM\cN}%^{\bf 3875}
  \;,
 \end{split}
 \eea
where $c_1$ and $c_2$ are two arbitrary constants.
 This allows to almost recover the
duality relation (\ref{vectordual}) from the $E_{10}$ equations of
motion (\ref{forenklat2}). First, (\ref{forenklat2}) may be rewritten as
 \bea
  \partial_t\left(n^{-1}g^{mn}\vG_{\cM\cN}\partial_tA_n{}^{\cN}+\ft12c_1g
  \varepsilon^{mn}\eta_{\cM\cN}\theta A_n{}^{\cN}
  +14c_2g\varepsilon^{mn}\tilde{\Theta}{}_{\cM\cN}%^{\bf 3875}
  A_n{}^{\cN}\right) \ = \ 0\;.
 \eea
Therefore, it can be integrated to the first-order equation
 \bea\label{firstorder2}
  n^{-1}g^{mn}\vG_{\cM\cN}\partial_tA_n{}^{\cN} \ = \
  g\varepsilon^{mn}\Theta_{\cM\cN}A_n{}^{\cN}+\Xi^m{}_{\cM}\;.
 \eea
Here we have chosen the free constants to be $c_1=2$ and $c_2=1/14$
in order to conveniently combine the irreducible parts of the
embedding tensor into $\Theta_{\cM\cN}$ according to
\cite{Nicolai:2001sv}. Moreover, $\Xi^m{}_{\cM}$ denotes an
integration constant. This integration constant cannot be set to
zero without breaking the symmetries. The situation is analogous to
the integration leading to the embedding tensor $\Theta_{\cM\cN}$ in
(\ref{embedd}), which generically breaks the global $E_{8}$ symmetry
once $\Theta_{\cM\cN}$ is constant. Correspondingly, the $E_{10}$
shift symmetry leaves this first-order equation only invariant if
the integration constant also transforms as a shift,
 \bea
  \delta_{\Lambda}\Xi^m{}_{\cM} \ = \
  -g\varepsilon^{mn}\Theta_{\cM\cN}\Lambda_n{}^{\cN}\;,
 \eea
which is consistent with the time-independence of $\Xi$. Thus,
fixing it to any specific value (as zero) breaks the symmetry, and
in this sense supergravity may at best be viewed as a broken phase
of $E_{10}$. After setting $\Xi=0$ and contracting with
$\Theta_{\cM\cN}$, (\ref{firstorder2}) implies
 \bea\label{e10dual}
   g\Theta_{\cM\cN}\varepsilon^{mn}\partial_{t}A_{n}{}^{\cN}
   +g^{2}eN\vG^{\cK\cL}g^{mn}\Theta_{\cM\cK}\Theta_{\cN\cL}
   A_{n}{}^{\cN} \ = \ 0\;,
 \eea
which coincides with the duality relation (\ref{vectordual}) from
supergravity up to the replacement $\vG^{\cM\cN}\rightarrow
\ft12(\vG^{\cM\cN}+\eta^{\cM\cN})$.

Finally, insertion of (\ref{embedd}) and (\ref{e10dual}) into the
equations of motion (\ref{e10zero}) for $g_{mn}$ and $G^{\cM\cN}$ as
obtained from $E_{10}$ yields
 \bea\label{finale10}
  \begin{split}
   \frac{\delta L_0}{\delta
   g^{mn}}&+\ft18g^{2}\tte G^{\cM\cN}\Theta_{\cM\cK}\Theta_{\cN\cL}
   \left(g_{mn}g^{kl}A_{k}{}^{\cK}A_{l}{}^{\cL}-A_{m}{}^{\cK}A_{n}{}^{\cL}\right)\\
      &+\ft12g^2\tte g_{mn}\left(\tfrac{1}{56}\vG^{\cM\cK}\vG^{\cN\cL}\tilde{\Theta}{}_{\cM\cN}%^{\bf 3875}
  \tilde{\Theta}{}_{\cK\cL}
   +\theta^2\right)
   \ = \
   0\;, \\
   \frac{\delta L_0}{\delta
   G^{\cM\cN}}&-\ft18g^2\tte g^{mn}\Theta_{\cM\cK}\Theta_{\cN\cL}A_{m}{}^{\cK}A_{n}{}^{\cL}
   -\tfrac{1}{56}g^2\tte \vG^{\cK\cL}\tilde{\Theta}{}_{\cM\cK}
   \tilde{\Theta}{}_{\cN\cL}
    \ = \ 0\;.
  \end{split}
 \eea
Here, we have used (\ref{lapse}) and (\ref{dict01}).
By comparing (\ref{finale10}) with (\ref{Einsteineom}) and
(\ref{scalareom}) we observe that the equations are structure-wise
the same, but differ in the details. For one thing, on the $E_{10}$
side we generically have just $\vG^{\cM\cN}$ instead of
$\ft12(\vG^{\cM\cN}+\eta^{\cM\cN})$. Apart from that, the indefinite
contributions to the supergravity potential are not reproduced, but
only the leading term quadratic in $\vG^{\cM\cN}$.

Let us now inspect the simultaneous inclusion of gauge couplings and
spatial gradients. As before this requires an analysis at the level of 
$\cP$ that carries flat indices. 
Specifically, we can 
supplement the dictionary (\ref{dict01}) and (\ref{dict1})
with 
\begin{align}\label{dict2}
P^{\vA\vB}(t) &\equiv \tfrac1{28} N  g\tilde{T}_{\vA\vB}(t,{\bf x}_0),  &
P(t) &\equiv  N g\theta(t,{\bf x}_0)
\end{align}
on level $\ell=2$. This dictionary is derived from the integrability
conditions (\ref{sugralev1IJ})--(\ref{sugralev1A}) such that they match
exactly the common terms in  the equations (\ref{e10lev1IJ})--(\ref{e10lev1A})
for $E_{10}$ (the terms involving $Q_a^{IJ}$ do not match just as in
the Einstein equation (\ref{ell1einst})).  
Moreover, we have `covariantized' the dictionary since it only fixes $P^{AB}$
and $P^{A\,IJ}$, but not $P^{IJ\,KL}$.
However, using the dictionary (\ref{dict2}) in the Einstein equation one
finds that the scalar potential is not reproduced correctly. The terms coming
from the positive definite $(A_2^{I\dot{A}})^2$ contribution in (\ref{apot}),
however, appear precisely in the $E_{10}$ Einstein equation. 
If we only consider the terms in the scalar equation of motion arising
from the positive definite part,
then the dictionary (including also $P^{IJKL}$) gives the 
correct relative coefficients, but the overall coefficient is wrong. This can
be seen by comparing (\ref{sugrascalarshalva}) and (\ref{e10scalars}). 
For the full
potential we find disagreement  
since the potential is not positive-definite, unlike the Cartan-Killing
form used on the $E_{10}$ side, and one can see that 
there is no choice for 
the dictionary such that all equations match.
In addition, it is not the case that $E_{10}$ predicts a different potential.
Rather, the scalar dependence in the $E_{10}$ equations is such that it
cannot be 
integrated to a corresponding single scalar potential in a $D=3$ field theory.
To summarize, while there is no precise agreement between the corresponding
equations, the $E_{10}$ model predicts and provides an embedding tensor in
the correct $E_8$ representation, which in the present truncation is forced to
be constant by the geodesic equations. It is noteworthy that the $E_{10}$
model naturally contains both the constant embedding tensor and the scalar
field dependent $T$-tensor via dressing with the level zero vielbein.

Finally, we comment on the meaning of the field $B_{mn}{}^{\cM}$,
which we truncated so far. One possible interpretation might be as a
spatial gradient. Another attractive scenario is that it is related
to a novel type of gauging, the so-called trombone gauging, which
has recently appeared in the literature \cite{Diffon:2008sh}. This
gauging gives rise to embedding tensor components $\Theta_{\cM}$,
and it has been noted that they are in one-to-one correspondence
with certain mixed Young tableaux representations within $E_{11}$
and $E_{10} $\cite{Diffon:2008sh}. Applied to $D=3$ these degenerate
to the symmetric $B_{mn}{}^{\cM}$ and so one might hope to interpret
this as a trombone gauging. However, given the ambiguity of the
possible interpretations encountered so far, we postpone a detailed
analysis of this proposal to future work.

\subsection{Quadratic and gauge constraints}

We now turn to a discussion of the constraint equations that supplement the
dynamical equations discussed so far. From the $E_{10}$ point of view these
have to be considered as additional constraints on the geodesic. In
\cite{Damour:2007dt} it has been shown 
that the constraint equations 
in maximal eleven-dimensional supergravity can be consistently imposed on the geodesic and are weakly conserved as the system evolves. Furthermore, the constraints there followed an intriguing pattern, displaying a certain grading property reminiscent of a Sugawara-type construction in terms of bilinear products of conserved currents. Here, we will encounter a similar phenomenon which extends up to the quadratic constraint, probing generators of $E_{10}$ beyond the analysis carried out in \cite{Damour:2007dt}.

Besides the Hamiltonian constraint, the constraint equations which have to be studied in the present context are
\begin{itemize}
\item[$(i)$]
the diffeomorphism constraint (the $(0a)$ component of the Einstein equation (\ref{einstein})),
\item[$(ii)$] the Gauss constraint (\ref{gauss}) or (\ref{constraint1}),
\item[$(iii)$] the spatial constancy of $\Theta_{\cM\cN}$ (\ref{spattheta}) and
\item[$(iv)$] the quadratic constraint (\ref{quadtheta}) of standard gauging and possibly trombone gauging.
\end{itemize}
The first one arises from gauge fixing the shift vector $N^a=0$, whereas the
other three are all consequences of adopting
the temporal gauges (\ref{tempgauge}) for the tensors of gauged
supergravity. There are no additional Bianchi type constraints as there were
for $D=11$ supergravity in~\cite{Damour:2007dt} since these vanish identically
in $D=3$. For example, the equation $D_{[a} \ttF_{bc]}{}^\cM=0$ is fulfilled
trivially since there are no three distinct spatial indices $a,b,c$.

Analyzing the four constraint equations with the use of the dictionaries
derived in (\ref{dict0}), (\ref{dict1}) and (\ref{embedd}), and using the duality relation (\ref{covdual}), one finds that they have the schematic form
\bea
\coC_a  &=& P_a{}^A P^A \;,\nn\\
\coC^\cA &=& P^{\cA B} P^B + f^{\cA}{}_{\cB\cC}\varepsilon^{ab} P_a{}^\cB P_b{}^\cC\;,\nn\\
\coC_a{}^{\cA\cB} &=& f^{(\cA}{}_{\cC\cD} P^{\cB)\cC} P_a{}^\cD \;,\nn\\
\coC^{\cA\cB,\cC} &=& P^{\cC\cD} P^{\cE(\cA} f_{\cD\cE}{}^{\cB)}\;,
\eea
in flat indices (where the $SO(16)$ spinor indices $A$ and $B$ should not be confused with the adjoint $E_8$ indices $\cA$ and $\cB$). 
The important feature of these equations is the tensor structure and the fact that the levels of the $P$ components occurring on the right hand side always add up to the same number in each constraint. In this way one can assign to the four equations the `levels' $\ell=1,2,3,4$, respectively, since in the first one the combinations are $P^{(0)} P^{(1)}$ up to $P^{(2)} P^{(2)}$ in the last equation. Furthermore, they transform (after conversion to curved indices) in the
$GL(2,\,\R)\times E_{8}$ representations indicated. As in \cite{Damour:2007dt}
we can thus bring the above constraints into a Sugawara-like form by switching
to curved indices $m,n,\dots$ and $\cM,\cN,\dots$, and by replacing 
the $P$'s by the corresponding components of the conserved
$E_{10}$ Noether current.

In \cite{Damour:2007dt} it was also noted that the representation content of 
the graded constraints is very similar to that of a specific highest weight
representation of $E_{10}$, sometimes called $L(\Lambda_1)$ as it is the
highest weight module with highest weight corresponding to the fundamental
weight of node~1 of the $E_{10}$ Dynkin diagram in figure \ref{e10dynkin1}. We
give the decomposition of this representation with respect to
$SL(2,\,\R)\times E_8$ at low levels  in table~\ref{constable}. From this
table we see that there is again agreement between the representations of the
constraints at low levels and the tensors contained in the $L(\Lambda_1)$
representation. At higher levels there appear extra representations, some of
which can probably be interpreted as recurrences (higher order gradients) of
the constraints encountered before but this explanation seems incomplete and
therefore we have partly left the interpretation open.
\begin{table}[h]
\begin{centering}
\begin{tabular}{|c|c|c|}
    \hline
    Level $\ell$  & $SL(2,\,\R) \times E_8$ representation           & Interpretation \\
    \hline
    \hline
    &  & \\ [-10pt]
    1   & $( {\bf 2}, {\bf 1})$       &
Diffeomorphism constraint \\
    \hline
    &  & \\ [-10pt]
    2   & $( {\bf 1}, {\bf 248})$                    & Gauss constraint \\
    &  & \\ [-10pt]
    \hline
    &  & \\ [-10pt]
    3   & $( {\bf 2}, {\bf 1})$    &
    Spatial constancy of $\theta$\\
    & $({\bf 2}, {\bf 3875})$ &
    Spatial constancy of ${\tilde \Theta}_{\cM\cN}$ \\ %$Y^{[ij]}_{\hspace{0.3em}(\cM\cN)}$ \\
        & $( {\bf 2}, {\bf 248})$                    &
       Spatial constancy of $\Theta_{\cM}$ (trombone)? \\ %$Z^{(ij)}{}_{\cM}$ \\ [2pt]
           &  & \\ [-10pt]
    \hline
    &  & \\ [-10pt]
    4 & $({\bf 1}, {\bf 147250})$ & Quadratic constraint\\
       & $({\bf 3}, {\bf 30380})$ & Quadratic constraint of trombone?\\
       & $({\bf 1}, {\bf 30380})$ & ?\\
       & $({\bf 3}, {\bf 3875})$ & Quadratic constraint of trombone?\\
       & $2\times ({\bf 1}, {\bf 3875})$ & Quadratic constraint\\
       & $2\times ({\bf 3}, {\bf 248})$ & Quadratic constraint of trombone?\\
       & $2\times ({\bf 1}, {\bf 248})$ & Recurrence of Gauss?\\
       & $ ({\bf 3}, {\bf 1})$ & Quadratic constraint of trombone?\\
       & $({\bf 1}, {\bf 1})$ & Recurrence of $\theta$?\\
    \hline
\end{tabular}
\caption{
    $SL(2,\,\R) \times E_8$ decomposition of $L(\Lambda_1)$ highest weight representation of $E_{10}$.
    \label{constable}}
\end{centering}
\end{table}

We note that it is to be expected that the constraints only form a
representation of a Borel subgroup $E_{10}^+\subset E_{10}$
rather than of the whole $E_{10}$ since explicit calculations of the
transformation of the diffeomorphism constraint show that it is not
annihilated by elements of the conjugate subgroup
$E_{10}^-$~\cite{Damour:2007dt}.\footnote{Here, the $\pm$ superscripts on
  $E_{10}$ should not be confused with further Kac--Moody extensions of
  $E_{10}$ but refer to Borel subgroups generated by positive and negative
  level generators, respectively.}

\section{Discussion and outlook}
In this paper we explored the $E_{10}$/supergravity correspondence
for the case of gauged supergravity. Apart from the inclusion of
spatial gradients and/or mass parameters discussed in the literature
so far, this provides additional insights into the interpretation of
part of the higher-level representations within $E_{10}$. As has
been found before, in general dimensions $D$ there are $(D-1)$-forms
whose representations coincide with those of consistent gaugings in
supergravity. Moreover, here we found that the quadratic constraint
of gauged supergravity belongs to the same highest weight representation
of $E_{10}$ as the diffeomorphism constraint (but, we repeat, the
constraints transform properly only under the Borel part $E_{10}^+$
of that representation). In contrast, in the
$E_{11}$ approach the $D$-form Lagrangian multiplier for this
constraint arises as one of the higher-level fields. While at a
purely kinematical level the Kac-Moody algebras $E_{10}$ and
$E_{11}$ therefore encode gauged supergravity, the sigma model
theory discussed in this paper allows, in addition, to check the
correspondence at the level of dynamics.

Most remarkably, we find that the equations of motion of gauged
supergravity (here for the example of three space-time dimensions)
adapted to a one-dimensional language can in part be matched to the
$E_{10}$ equations, even though the latter have a priori a rather
different form. For one thing, the absence of gauge-covariant
derivatives on the $E_{10}$ side agrees with the supergravity
expressions, once the gauge-fixing condition $\ttA_{t}{}^{\cM}=0$,
which is inevitable for the comparison, has been imposed. Moreover,
in spite of the fact that on the $E_{10}$ side all fields appear
with a `kinetic' term, the (truncated) duality relation between
vectors and scalars expected from supergravity naturally follows via
integrating the one-dimensional equations of motion. Finally, the
embedding tensor automatically appears as an integration constant in
the right representation. In this sense, none of the essential
ingredients of gauged supergravity have to be introduced by hand,
but rather they naturally follow from the $E_{10}$ sigma model.

Irrespective of these promising observations, there remain mismatches
at higher levels, which prohibit a full agreement between supergravity and 
the $E_{10}$ model.
One finds 
systematically that
while in supergravity the combination $\ttG_{\cM\cN}+\eta_{\cM\cN}$
appears, the corresponding equations on the $E_{10}$ side only
contain $\vG_{\cM\cN}$. Similarly, the scalar potential is not fully
reproduced by $E_{10}$. This is due to the fact that in supergravity
the scalar potential is indefinite \cite{deWit:2008ta}, 
while the corresponding 2-forms
appearing in the $E_{10}$ coset model necessarily enter with a
positive-definite kinetic term. The latter is
somewhat reminiscent to a discrepancy encountered in higher
dimensions, once spatial gradients are introduced as the duals of
higher-level fields.

In total we are led to conclude that further insights are required
in order to understand the precise relation between supergravity
theories and the $E_{10}$ sigma model. It would be interesting to
see whether modifications and/or extensions of the $E_{10}$ model
are possible to compensate for the present mismatches. We note that
mismatches already occur before comparing to \textit{gauged}
supergravity and so an ultimate resolution of the present
discrepancies must await a better understanding of the basic
picture.

\bigskip

\noindent
\textbf{Acknowledgments}

The authors are grateful to Ella Jamsin, Marc Henneaux, Daniel Persson and
Henning Samtleben for interesting 
discussions and thank each other's home institutions for the hospitality
extended during various visits.
This work was partially supported by
the European Commission FP6 program MRTN-CT-2004-005104EU and by the
INTAS Project 1000008-7928. A.K. is a Research Associate of the Fonds de la
Recherche Scientifique--FNRS, Belgium. This work is part of the
research programme of the 'Stichting voor Fundamenteel Onderzoek der
Materie (FOM)', which is financially supported by the 'Nederlandse
Organisatie voor Wetenschappelijk Onderzoek (NWO)'.

%\newpage

\begin{appendix}
\renewcommand{\theequation}{\Alph{section}.\arabic{equation}}

\section{Conventions for $E_{8}$} \label{e8app}
Our conventions for $E_{8}$ are as in \cite{Nicolai:2001sv}. The
Lie algebra $\frak{e}_{8}$ is generated by $t^{\cM}$, with
$\cM,\cN,\ldots = 1,\ldots,248$ denoting the adjoint indices, and
bracket $[t^{\cM},t^{\cN}]=f^{\cM\cN}{}_{\cK}t^{\cK}$. Specifically,
$\frake_{8}$ can be defined according to its $\so(16)$ decomposition,
 \begin{eqnarray}\label{E8decom}
  [t^{IJ},t^{KL}]  &=& 4 \delta^{JK}t^{IL}
  %-
  %\delta^{IK}X^{JL}-\delta^{JL}X^{IK}+\delta^{IL}X^{JK}
  \;, \\
  \nonumber
  [t^{IJ},t^{A}] &=& -\tfrac{1}{2}\Gamma^{IJ}{}_{AB}t^{B}\;,  \\
  \nonumber
  [t^{A},t^{B}] &=& \tfrac{1}{4}\Gamma^{IJ}{}_{AB}t^{IJ}\;.
 \end{eqnarray}
Here $I,J,\ldots=1,\ldots,16$ are $SO(16)$ vector indices, while
$A,B,\ldots = 1,\ldots,128$ label spinor indices. The adjoint
indices split according to $\cA=([IJ],A)$, where we employ the
convention that summation over the antisymmetric $[IJ]$ is
accompanied by a factor of $\ft12$. The spinor generators are
defined by
$\Gamma^{I}{}_{A\dot{A}}\Gamma^{J}{}_{B\dot{A}}=\delta^{IJ}\delta_{AB}+\Gamma^{IJ}{}_{AB}$.
Like any other Kac-Moody algebra, $\mathfrak{e}_{8}$ admits an invariant Cartan-Killing form, which in
the $SO(16)$ decomposition (\ref{E8decom}) reads
 \bea\label{Cartan-Killing}
  \eta^{AB} \ = \ \delta^{AB}\;, \qquad
  \eta^{IJ\,KL} \ = \ -2\delta^{IK}\delta^{JL}.
 \eea
Accordingly, in the totally antisymmetric structure constants
 \bea
 \begin{split}
  f^{IJ\,KL\,MN} & \ = \  -f^{IJ\,KL}{}_{MN} \ = \ 8\delta^{IK}\delta^{J}{}_{M}\delta^{L}{}_{N}
  \\
  f^{IJ\,A\,B}& \ = \ -f_{IJ}{}^{A\,B} \ = \ -\tfrac{1}{2}\Gamma^{IJ}{}_{AB}
 \end{split}
 \eea
we can freely raise and lower indices. We recall that we use the convention
that the right hand side is always to be antisymmetrized in the same way as
the left hand side.
The $E_8$ structure constants and the Killing form
are related by the identity
$f^{\vA\vB}{}_{\vC}f_{\vA\vB\vD}=-60\eta_{\vC\vD}$,
which implies
$f^{\vA\vB\vC}f_{\vA\vB\vC}=-14880$.
We also frequently use the
relation
 \bea
  \ttE^{-1}t^{\cM}\ttE \ = \ \ttE^{\cM}{}_{\cA}t^{\cA}\;
 \eea
for the adjoint matrix $\ttE\in E_8$, which can be easily checked by
use of the Baker-Campbell-Hausdorff formula (\ref{BCH}).

The
tensor product of two
adjoint representations
decomposes as
\begin{align}
\bf 248 \times 248 = 1+248+3875+27000+30380,
\end{align}
and the corresponding projectors have the components \cite{Koepsell:1999uj}
\begin{align} \label{e8proj}
(\mathbb{P}_{\bf1})_{\vA\vB}{}^{\vC\vD} &= \tfrac{1}{248}\eta_{\vA\vB}\eta^{\vC\vD},\nn\\
(\mathbb{P}_{\bf248})_{\vA\vB}{}^{\vC\vD} &= -\tfrac{1}{60}
f^{\vE}{}_{\vA\vB}f_{\vE}^{\vC\vD},\nn\\
(\mathbb{P}_{\bf3875})_{\vA\vB}{}^{\vC\vD} &= \tfrac{1}{7}\delta_{(\vA}{}^{\vC}\delta_{\vB)}{}^{\vD}
-\tfrac{1}{56}\eta_{\vA\vB}\eta^{\vC\vD}-\tfrac{1}{14}f^{\vE}{}_{\vA}{}^{(\vC}f_{\vE\vB}{}^{\vD)},\nn\\
(\mathbb{P}_{\bf27000})_{\vA\vB}{}^{\vC\vD} &= \tfrac{6}{7}\delta_{(\vA}{}^{\vC}\delta_{\vB)}{}^{\vD}
+\tfrac{3}{217}\eta_{\vA\vB}\eta^{\vC\vD}+\tfrac{1}{14}f^{\vE}{}_{\vA}{}^{(\vC}f_{\vE\vB}{}^{\vD)},\nn\\
(\mathbb{P}_{\bf30380})_{\vA\vB}{}^{\vC\vD} &=
\delta_{[\vA}{}^{\vC}\delta_{\vB]}{}^{\vD}+\tfrac{1}{60}f^{\vE}{}_{\vA\vB}f_{\vE}^{\vC\vD}.
\end{align}
Elsewhere in the paper, we have dropped the subscript on $\mathbb{P}_{\bf3875}$.
Splitting the indices, we get the following identities for a tensor $\tilde{T}^{\vA\vB}$
that transforms in the $\bf 3875$ representation:
\begin{align} \label{enklarevillkor}
\tilde{T}^{A\,IJ}&=-\tfrac16\Ga^{IK}{}_{AB}\tilde{T}^{B\,JK}
=\tfrac1{26}\Ga^{IJKL}{}_{AB}\tilde{T}^{B\,KL},\nn\\
\tilde{T}^{IJ\,KL} &= \tfrac37\de^{IK}\tilde{T}^{JM\,LM}-\tilde{T}^{IK\,JL},\nn\\
\tilde{T}^{AB}&=\tfrac1{96}
\Ga^{IJKL}{}_{AB}\tilde{T}^{IJ\,KL}.
\end{align}
The two equations in the first line are equivalent. The last equation can be inverted to
\begin{align}
\Ga^{IJKL}{}_{AB}\tilde{T}^{AB}=32\,\tilde{T}^{[IJ\,KL]}.
\end{align}
We also note that
$\tilde{T}^{IJ\,IJ}=\tilde{T}^{AA}=0$.

\section{Level decomposition of $E_{10}$} \label{e10app}
To determine the $E_{10}$ commutation relation
(\ref{postulat3}), we needed to
identify the
Chevalley generators, which are the 30 elements $h_i,\,e_i,\,f_i$ ($i=1,\,2,\,\ldots,\,10$) that satisy the Chevalley-Serre relations (\ref{chevrel}) and (\ref{serrerel}).
We let any $x \in \frake_8$ have the components $x_\vA$ in the $t^\vA$ basis,
$x = x_\vA t^\vA$. Then we get
\begin{align} \label{identification1}
e_1&=K^1{}_2, &
e_2&=(-f_\theta)_\vA E^{2}{}_{\vB}\eta^{\vA\vB},&
e_i&=(e_i)_\vA t^{\vA},\nn\\
h_1&=K^1{}_1-K^2{}_2,&
h_2&=(-h_\theta)_\vA t^{\vA} -K^1{}_1,&
h_i&=(h_i)_\vA t^{\vA},\nn\\
f_1&=K^2{}_1, &
f_2&=(-e_\theta)_\vA F_2{}^\vA,&
f_i&=(f_i)_\vA t^{\vA},
\end{align}
for $i=3,\,4,\,\ldots,\,10$. 
Here $\theta$ (not to be confused with the singlet embedding tensor) denotes the highest root of $\frake_8$, with the corresponding step operators $e_\theta,\,f_\theta$ and Cartan element
$h_\theta$.
We have
\begin{align}
h_{\theta}=2h_3+3h_4+4h_5+5h_6+6h_7+4h_8+2h_9+3h_{10}
\end{align}
and we get
\begin{align}
K^1{}_1&= -h_\theta-h_2, & K^2{}_2&= -h_\theta-h_2-h_1, &
K&=-2h_\theta-2h_2-h_1.
\end{align}
By inserting (\ref{identification1}) into (\ref{chevrel}) and using
(\ref{liealgebra}), we see that the Chevalley relations 
 $[h_i, e_j]  = A_{ij} e_j$ and $[h_i, f_j]  = -A_{ij} f_j$ are indeed satisfied.
 For the remaining relations to hold, $[e_i, f_j]  = \delta_{ij} h_i$, we must have
\begin{align}
[E^{a}{}_{\vA},\,F_{b}{}^{\vB}]&=\delta^a{}_bf_{\vA}{}^{\vB}{}_{\vC}t^{\vC}
+\delta_{\vA}{}^{\vB}K^a{}_b-\delta_{\vA}{}^{\vB}\delta^a{}_b K, 
\end{align}
where we have set
$K=K^a{}_a=K^1{}_1+K^2{}_2$.
The relations (\ref{postulat1}) and (\ref{postulat3}) can then be inverted to
\begin{align}
E&=\tfrac{1}{248}\varepsilon_{ab}\eta^{\vA\vB}
[E^{a}{}_{\vA},\,E^{b}{}_{\vB}],\nn\\
E_{\vA \vB}&=
\tfrac{1}{2}\varepsilon_{ab}[E^{a}{}_{\vA},\,E^{b}{}_{\vB}]
-\tfrac{1}{496}\varepsilon_{ab}\eta_{\vA\vB}\eta^{\vC\vD}[E^{a}{}_{\vC},\,E^{b}{}_{\vD}],\nn\\
E^{ab}{}_{\vA}&=\tfrac{1}{60}f_\vA{}^{\vB\vC}[E^{a}{}_{\vB},\,E^{b}{}_{\vC}],%\nn
\end{align}
\begin{align}
F&=-\tfrac{1}{248}\varepsilon^{ab}%\eta_{\vA\vB}[
\eta_{\vA\vB}[F_{a}{}^{\vA},\,F_{b}{}^\vB],\nn\\
F^{\vA \vB}&=
-\tfrac{1}{2}\varepsilon^{ab}[F_{a}{}^\vA,\,F_{b}{}^\vB]
+\tfrac{1}{496}\varepsilon^{ab}\eta^{\vA\vB}\eta_{\vC\vD}
[F_{a}{}^\vC,\,F_{b}{}^\vD],\nn\\
F_{ab}{}^\vA&=\tfrac{1}{60}f^\vA{}_{\vB\vC}[F_{a}{}^\vB,\,F_{b}{}^{\vC}],%\nn
\end{align}
\begin{align}
t^{\vA}&=-\tfrac{1}{120}f^{\vA\vB}{}_{\vC}[E^{a}{}_{\vB},\,F_{a}{}^\vC],&
K^a{}_b&=\tfrac{1}{248}([E^{a}{}_{\vA},\,F_{b}{}^{\vA}]-
\delta^a{}_b[E^{c}{}_{\vA},\,F_{c}{}^{\vA}]).
\end{align}
The remaining nonzero commutation relations follow from the Jacobi identity,
\begin{align}
[E,\,F_a{}^{\vA}]&=-\tfrac{1}{2}\varepsilon_{ab}\eta^{\vA\vB}E^{b}{}_{\vB}, &
[F,\,E^a{}_{\vA}]&=\tfrac{1}{2}\varepsilon^{ab}\eta_{\vA\vB}F_{b}{}^{\vB},\nn\\
[E^{ab}{}_\vA,\,F_c{}^\vB]&=
-\delta^a{}_c f_{\vA}{}^{\vB\vC} E^{b}{}_{\vC},&
[F_{ab}{}^\vA,\,E^c{}_\vB]&=
-\delta^c{}_a f^{\vA}{}_{\vB\vC}F_{b}{}^{\vC},
\nn\\
[E_{\vA\vB},\,F_a{}^\vC]&=-14\ep_{ab}\mathbb{P}_{\vA\vB}{}^{\vC\vD} E^{b}{}_{\vD}, &
[F^{\vA\vB},\,E^a{}_\vC]&=14\ep^{ab}\mathbb{P}^{\vA\vB}{}_{\vC\vD} F_b{}^{\vD},%\nn
\end{align}
\begin{align}
[E_{\vA\vB},\,F^{\vC\vD}]&=
\label{abcd-kommutator}
2f^{\vC}{}_{\vA}{}_{\vE}f^{\vE\vD}{}_{\vF}f^{\vF}{}_{\vB}{}_{\vG}t^{\vG}
-4\delta_{\vA}{}^{\vC}f^{\vD}{}_{\vB}{}_{\vE}t^{\vE}%\\&\quad
-14{\mathbb{P}}_{\vA\vB}{}^{\vC\vD}K,\nn
%[E,\,F]&=-K,
\end{align}
\begin{align}
[E^{ab}{}_{\vA},\,F_{cd}{}^{\vB}]&=f_{\vA}{}^{\vB}{}_{\vC}t^\vC
+2\delta_{\vA}{}^{\vB}(
\delta^{a}{}_{c}K^{b}{}_{d}-%2\eta^{\vA\vB}
\delta^{a}{}_{c}\delta^{b}{}_{d}K), & [E,\,F]&=-K.
\end{align}
Here we have used that $\ep^{ac}\ep_{cb}=-\de^a{}_b$ with our conventions.
Using the invariance of the Cartan-Killing form, we have
\begin{align}
-\tfrac{1}{4}\la [E^{a}{}_{\vA},\,E^{b}{}_{\vB}] | [F_{c}{}^{\vC},\,F_{d}{}^{\vD}] \ra
=(31
\mathbb{P}_{({\bf 1},\,{\bf 1})}
+15 
\mathbb{P}_{({\bf 3},\,{\bf 248})}
+7 
\mathbb{P}_{({\bf 1},\,{\bf 3875})}){}^{ab}{}_{cd}{}^{\vC\vD}{}_{\vA\vB},
\end{align}
where $\mathbb{P}_{({\bf 1},\,{\bf 1})},\,
\mathbb{P}_{({\bf 3},\,{\bf 248})}$ and
$\mathbb{P}_{({\bf 1},\,{\bf 3875})}$
are the projectors corresponding to the
$SL(2,\,\R) \times E_8$ representations at level $\ell=2$ (cf.~Table \ref{reptable}). Explicitly,
\begin{align}
\mathbb{P}_{({\bf 1},\,{\bf 1})}{}^{ab}{}_{cd}{}^{\vC\vD}{}_{\vA\vB}&=
\delta^{a}{}_{[c}\delta^{b}{}_{d]}
\mathbb{P}_{{\bf 1}}{}^{\vC\vD}{}_{\vA\vB},\nn\\
\mathbb{P}_{({\bf 3},\,{\bf 248})}{}^{ab}{}_{cd}{}^{\vC\vD}{}_{\vA\vB}&=
\delta^{a}{}_{(c}\delta^{b}{}_{d)}
\mathbb{P}_{{\bf 248}}{}^{\vC\vD}{}_{\vA\vB},\nn\\
\mathbb{P}_{({\bf 1},\,{\bf 3875})}{}^{ab}{}_{cd}{}^{\vC\vD}{}_{\vA\vB}&=
\delta^{a}{}_{[c}\delta^{b}{}_{d]}
\mathbb{P}_{{\bf 3875}}{}^{\vC\vD}{}_{\vA\vB},
\end{align}
where the $E_8$ projectors $\mathbb{P}_{{\bf 1}},\,\mathbb{P}_{{\bf 248}}$
and $\mathbb{P}_{{\bf 3875}}$ \cite{Koepsell:1999uj} were already given in (\ref{e8proj}).

\end{appendix}


\begin{thebibliography}{99}

%\cite{Nahm:1977tg}
\bibitem{Nahm:1977tg}
  W.~Nahm,
  {\it Supersymmetries and their representations},
  Nucl.\ Phys.\  B {\bf 135} (1978) 149.
  %%CITATION = NUPHA,B135,149;%%

%\cite{Cremmer:1978km}
\bibitem{Cremmer:1978km}
  E.~Cremmer, B.~Julia and J.~Scherk,
  {\it Supergravity theory in 11 dimensions,}
  Phys.\ Lett.\  B {\bf 76} (1978) 409.
  %%CITATION = PHLTA,B76,409;%%

\bibitem{1979NuPhB.159..141C}
E.~Cremmer and B.~Julia,
{\it The $SO(8)$ supergravity},
{Nucl.\ Phys.\ B}
{\bf159} (1979) {141}.%--212}.

\bibitem{1983uft..conf..215J}
B.~{Julia},  {\em {Application of supergravity to gravitation theory}}, in {\em
  Unified field theories of $>4$ dimensions}, pp.~215--236.
\newblock World Scientific, 1983.

\bibitem{Marcus:1983hb}
N.~Marcus and J.~H. Schwarz,  {\em {Three-dimensional supergravity theories}},
  Nucl. Phys. {\bf B228}, 145
(1983).
%%CITATION = NUPHA,B228,145;%%.

\bibitem{Nicolai:2000sc}
  H.~Nicolai and H.~Samtleben,
  {\it Maximal gauged supergravity in three dimensions},
  Phys.\ Rev.\ Lett.\  {\bf 86} (2001) 1686
  {\tt[arXiv:hep-th/0010076]}.

\bibitem{Nicolai:2001sv}
  H.~Nicolai and H.~Samtleben,
  {\it Compact and noncompact gauged maximal supergravities in three
  dimensions},
  JHEP {\bf 0104} (2001) 022
  {\tt[arXiv:hep-th/0103032]}.

\bibitem{Julia:1982gx}
  B.~Julia,
  {\it Kac-Moody Symmetry Of Gravitation And Supergravity Theories}, in:
  M. Flato, P. Sally and G. Zuckerman (eds.), 
  Applications of Group Theory in Physics and Mathematical Physics
  (Lectures in Applied Mathematics 21), Am. Math. Soc. (Providence,
  1985) 355--374, LPTENS 82/22
  %%CITATION = C82-07-06;%%

%\cite{West:2001as}
\bibitem{West:2001as}
  P.~C.~West,
  {\it $E_{11}$ and M theory},
  Class.\ Quant.\ Grav.\  {\bf 18} (2001) 4443
  {\tt[arXiv:hep-th/0104081]}.
  %%CITATION = CQGRD,18,4443;%%


%\cite{Schnakenburg:2001ya}
\bibitem{Schnakenburg:2001ya}
  I.~Schnakenburg and P.~C.~West,
  {\it Kac-Moody symmetries of IIB supergravity},
  Phys.\ Lett.\  B {\bf 517} (2001) 421
  {\tt[arXiv:hep-th/0107181]}.
  %%CITATION = PHLTA,B517,421;%%


%\cite{Kleinschmidt:2003mf}
\bibitem{Kleinschmidt:2003mf}
  A.~Kleinschmidt, I.~Schnakenburg and P.~West,
  {\it Very-extended Kac--Moody algebras and their interpretation at low
    levels}, 
  Class.\ Quant.\ Grav.\  {\bf 21} (2004) 2493
  {\tt[arXiv:hep-th/0309198]}.
  %%CITATION = CQGRD,21,2493;%%

\bibitem{Damour:2002cu}
  T.~Damour, M.~Henneaux and H.~Nicolai,
  {\it $E_{10}$ and a 'small tension expansion' of M-theory},
  Phys.\ Rev.\ Lett.\  {\bf 89} (2002) 221601
  {\tt[arXiv:hep-th/0207267]}.

%\cite{Damour:2004zy}
\bibitem{Damour:2004zy}
  T.~Damour and H.~Nicolai,
  {\it Eleven dimensional supergravity and the $E_{10}$/$K(E_{10})$
    sigma-model at  low  $A_{9}$ levels}, in: G.~S.~Pogoyan, L.~E.~Vicent and
  K.~B.~Wolf (eds.), Group 
  Theoretical Methods in Physics (IoP Conference Series Number 185), IoP
  Publishing  (2005) 93--111 
  {\tt[arXiv:hep-th/0410245]}.
  %%CITATION = HEP-TH/0410245;%%

\bibitem{Damour:2007dt}
  T.~Damour, A.~Kleinschmidt and H.~Nicolai,
  {\it Constraints and the $E_{10}$ coset model},
  Class.\ Quant.\ Grav.\  {\bf 24} (2007) 6097
  {\tt[arXiv:0709.2691 [hep-th]]}.
  %%CITATION = CQGRD,24,6097;%%

\bibitem{Damour:2005zs}
  T.~Damour, A.~Kleinschmidt and H.~Nicolai,
  {\it Hidden symmetries and the fermionic sector of eleven-dimensional
  supergravity},
  Phys.\ Lett.\  B {\bf 634} (2006) 319
  {\tt[arXiv:hep-th/0512163]}.

\bibitem{deBuyl:2005mt}
  S.~de Buyl, M.~Henneaux and L.~Paulot,
  {\it Extended $E_8$ invariance of 11-dimensional supergravity},
  JHEP {\bf 0602} (2006) 056
  {\tt[arXiv:hep-th/0512292]}.

\bibitem{Damour:2006xu}
  T.~Damour, A.~Kleinschmidt and H.~Nicolai,
  {\it $K(E_{10})$, supergravity and fermions},
  JHEP {\bf 0608} (2006) 046
  {\tt[arXiv:hep-th/0606105]}.

\bibitem{Englert:2003py}
  F.~Englert and L.~Houart,
  {\it ${\cal{G}}^{+++}$ invariant formulation of gravity and M-theories: Exact
    BPS   solutions},
  JHEP {\bf 0401} (2004) 002
  [arXiv:hep-th/0311255].
  %%CITATION = JHEPA,0401,002;%%

\bibitem{Englert:2004it}
  F.~Englert and L.~Houart,
  {\it ${\cal G}^{+++}$ invariant formulation of gravity and M-theories: Exact
      intersecting brane solutions},
  JHEP {\bf 0405} (2004) 059
  [arXiv:hep-th/0405082].
  %%CITATION = JHEPA,0405,059;%%

%\cite{Nicolai:2003fw}
\bibitem{Nicolai:2003fw}
  H.~Nicolai and T.~Fischbacher,
 {\it Low level representations for $E_{10}$ and $E_{11}$}, Contribution to
 the Proceedings of the Ramanujan International Symposium on 
  Kac--Moody Algebras and Applications, ISKMAA-2002, Chennai, India
  {\tt[arXiv:hep-th/0301017]}.
  %%CITATION = HEP-TH/0301017;%%


 \bibitem{Riccioni:2006az}
  F.~Riccioni and P.~West,
  {\it Dual fields and $E_{11}$},
  Phys.\ Lett.\  B {\bf 645} (2007) 286
  {\tt[arXiv:hep-th/0612001]}.


  %\cite{Kleinschmidt:2004dy}
\bibitem{Kleinschmidt:2004dy}
  A.~Kleinschmidt and H.~Nicolai,
  {\it $E(10)$ and $SO(9,9)$ invariant supergravity},
  JHEP {\bf 0407} (2004) 041
  {\tt[arXiv:hep-th/0407101]}.
  %%CITATION = JHEPA,0407,041;%%

%\cite{Romans:1985tz}
\bibitem{Romans:1985tz}
  L.~J.~Romans,
  {\it Massive N=2A supergravity in ten dimensions},
  Phys.\ Lett.\  B {\bf 169} (1986) 374.
  %%CITATION = PHLTA,B169,374;%%

\bibitem{Henneaux} M.~Henneaux, E.~Jamsin, A.~Kleinschmidt and D.~Persson,
  {\it On the $E_{10}$/massive IIA correspondence}, {\tt arXiv:0811.4358
    [hep-th]}. 

  %\cite{Schnakenburg:2002xx}
\bibitem{Schnakenburg:2002xx}
  I.~Schnakenburg and P.~C.~West,
  {\it Massive IIA supergravity as a non-linear realisation},
  Phys.\ Lett.\  B {\bf 540} (2002) 137
  {\tt[arXiv:hep-th/0204207]}.
  %%CITATION = PHLTA,B540,137;%%

%\cite{Riccioni:2007au}
\bibitem{Riccioni:2007au}
  F.~Riccioni and P.~C.~West,
  {\it The $E_{11}$ origin of all maximal supergravities},
  JHEP {\bf 0707} (2007) 063
  {\tt [arXiv:0705.0752 [hep-th]]}.
  %%CITATION = JHEPA,0707,063;%%

\bibitem{Bergshoeff:2007qi}
  E.~A.~Bergshoeff, I.~De Baetselier and T.~A.~Nutma,
  {\it $E_{11}$ and the embedding tensor},
  JHEP {\bf 0709} (2007) 047
  {\tt[arXiv:0705.1304 [hep-th]]}.


%\cite{Bergshoeff:2007vb}
\bibitem{Bergshoeff:2007vb}
  E.A.~Bergshoeff, J.~Gomis, T.A.~Nutma and D.~Roest,
  {\it Kac-Moody spectrum of (half-) maximal supergravities},
  JHEP {\bf 0802} (2008) 069
  {\tt [arXiv:0711.2035~[hep-th]]}.
  %%CITATION = JHEPA,0802,069;%%

%\cite{Riccioni:2007ni}
\bibitem{Riccioni:2007ni}
  F.~Riccioni and P.~C.~West,
  {\it $E_{11}$-extended spacetime and gauged supergravities},
  JHEP {\bf 0802} (2008) 039
  {\tt [arXiv:0712.1795 [hep-th]]}.
  %%CITATION = JHEPA,0802,039;%%

\bibitem{SimpLie}
    SimpLie: a simple program for Lie algebras,\\{\tt http://strings.fmns.rug.nl/SimpLie/}.

    \bibitem{Koepsell:1999uj}
  K.~Koepsell, H.~Nicolai and H.~Samtleben,
  {\it On the Yangian $[Y(\mathfrak{e}_{8})]$ quantum symmetry of maximal supergravity in  two
  dimensions},
  JHEP {\bf 9904} (1999) 023
  {\tt[arXiv:hep-th/9903111]}.

\bibitem{Bergshoeff:2008qd}
  E.~A.~Bergshoeff, O.~Hohm and T.~A.~Nutma,
  {\it A note on $E_{11}$ and three-dimensional gauged supergravity},
  JHEP {\bf 0805} (2008) 081
  {\tt[arXiv:0803.2989 [hep-th]]}.
  %%CITATION = JHEPA,0805,081;%%


\bibitem{deWit:2008ta}
  B.~de Wit, H.~Nicolai and H.~Samtleben,
  {\it Gauged Supergravities, Tensor Hierarchies, and M-Theory},
  JHEP {\bf 0802} (2008) 044
  {\tt [arXiv:0801.1294 [hep-th]]}.
  %%CITATION = JHEPA,0802,044;%%

%\cite{deWit:2005hv}
\bibitem{deWit:2005hv}
  B.~de Wit and H.~Samtleben,
{\it Gauged maximal supergravities and hierarchies of nonabelian
 vector-tensor  systems},
  Fortsch.\ Phys.\  {\bf 53} (2005) 442
  {\tt[arXiv:hep-th/0501243]}.
  %%CITATION = FPYKA,53,442;%%



\bibitem{Kleinschmidt:2005bq}
  A.~Kleinschmidt and H.~Nicolai,
  {\it Gradient representations and affine structures in $AE_{n}$},
  Class.\ Quant.\ Grav.\  {\bf 22} (2005) 4457
  {\tt[arXiv:hep-th/0506238]}.
  %%CITATION = CQGRD,22,4457;%%

\bibitem{Diffon:2008sh}
  A.~Le Diffon and H.~Samtleben,
  {\it Supergravities without an action: gauging the trombone},
  {\tt arXiv:0809.5180 [hep-th]}.




\end{thebibliography}
\end{document}